# Optical observations of "hot" novae returning to quiescence

P. Zemko[1⋆], S. Ciroi[1], M. Orio[2,3], A. Odendaal[4], S. Shugarov[5,6], E. Barsukova[7], A. Bianchini[3], V. Cracco[1] M. Gabdeev[7], V. Goranskij[5], B. Tofflemire[2], A. F. Valeev[7,8], N. Katisheva[9]

[1] *Department of Physics and Astronomy, Università di Padova, vicolo dell'Osservatorio 3, I-35122 Padova, Italy*
[2] *Department of Astronomy, University of Wisconsin, 475 N. Charter Str., Madison, WI 53704, USA*
[3] *INAF - Osservatorio di Padova, vicolo dell' Osservatorio 5, I-35122 Padova, Italy*
[4] *Department of Physics, University of the Free State, 9300, Bloemfontein, South Africa*
[5] *Sternberg Astrophysical Institute, Moscow University, Universitetsky Ave., 13, Moscow 119992, Russia*
[6] *Astronomical Institute of the Slovak Academy of Sciences, Tatranská Lomnica, 059 60, The Slovak Republic*
[7] *Special Astrophysical Observatory, Russian Academy of Sciences, Nizhnij Arkhyz, Karachai-Cherkesia, 369167, Russia*
[8] *Kazan Federal University, Kremlevskaya 18, 420008 Kazan, Russia*
[9] *Sternberg Astronomical Institute, Lomonosov Moscow University, 13, University av. Moscow, 119992, Russia*

**ABSTRACT**

We have monitored the return to quiescence of novae previously observed in outburst as supersoft X-ray sources, with optical photometry of the intermediate polar (IP) V4743 Sgr and candidate IP V2491 Cyg, and optical spectroscopy of these two and seven other systems. Our sample includes classical and recurrent novae, short period (few hours), intermediate period (1-2 days) and long period (symbiotic) binaries. The light curves of V4743 Sgr and V2491 Cyg present clear periodic modulations. For V4743 Sgr, the modulation occurs with the beat of the rotational and orbital periods. If the period measured for V2491 Cyg is also the beat of these two periods, the orbital one should be almost 17 hours. The recurrent nova T Pyx already shows fragmentation of the nebular shell less than 3 years after the outburst. While this nova still had strong [O III] at this post-outburst epoch, these lines had already faded after 3 to 7 years in all the others. We did not find any difference in the ratio of equivalent widths of high ionization/excitation lines to that of the H$\beta$ line in novae with short and long orbital period, indicating that irradiation does not trigger high mass transfer rate $\dot{m}$ from secondaries with small orbital separation. An important difference between the spectra of RS Oph and V3890 Sgr and those of many symbiotic persistent supersoft sources is the absence of forbidden coronal lines. With the X-rays turn-off, we interpret this as an indication that mass transfer in symbiotics recurrent novae is intermittent.

**Key words:** line: profiles – stars: novae, cataclysmic variables, individual (HV Cet, V2491 Cyg, KT Eri, RS Oph, T Pyx, U Sco, V3890 Sgr, V4743 Sgr, V382 Vel), white dwarfs, binaries: symbiotic

## 1 INTRODUCTION

Classical novae (CNe) are white dwarf (WD) binaries in which a high amount of accreted hydrogen-rich material resulted in a thermonuclear runaway from the surface of the white dwarf (WD). Although there is evidence that all novae have repeated outbursts, we call recurrent novae (RN) those with more than one nova explosion observed over human life timescales. Most RN should host massive WDs accreting at a high rate (see, for instance, models by Wolf et al. 2013). In this paper we present post-nova-explosion follow-up of binaries with two types of secondaries: late-type main sequence stars (cataclysmic variables) and red giants (symbiotic stars). The observations were done in the framework of a long term project of monitoring novae that were well observed in outburst in optical and X-rays, as they return to quiescence. Previous papers of this series include Zemko et al. (2015, hereafter Paper I) and Zemko et al. (2016, hereafter Paper II).

Novae are detected as bright sources in the optical band, but as the WD atmosphere contracts at rising temperature, the peak luminosity shifts towards shorter wavelengths. When the WD radius shrinks back to close to the pre-outburst size, the WD atmosphere heated by the underlying hydrogen burning, usually has temperature high enough to peak in the supersoft X-ray range. If the ejected envelope is transparent to soft X-rays, the system is observed as a super-soft X-ray source (SSS) (e.g. Orio et al. 2001a; Krautter 2008; Orio 2012). X-ray gratings observations of novae in the SSS phase allow to measure the chemical composition, the temperature of the WD atmosphere and the effective gravity. The last two parameters give an estimate of the WD mass. The SSS phase can last from days to years and ends when the WD atmosphere cools (Orio 2012).

⋆ E-mail:polina.zemko@studenti.unipd.it

In Papers I and II we continued to observe two recent novae that appeared as luminous SSS, V2491 Cyg (in X-rays) and V4743 Sgr (in X-rays and optical wavelengths), and found two interesting facts: first, both novae show diagnostics of intermediate polars (IPs), in which the accretion disk is disrupted by a moderately strong magnetic field of the WD (of the order of $10^6$ Gauss); second, in both cases the WD did not seem to cool isotropically, but the flux decreased by orders of magnitude with the effective temperature apparently remaining high in a small region. We hypothesized that this region may reside at the poles, and be due to the magnetic nature of the WD. We show in this paper that the optical light curves of these two novae show further evidence of their IP nature.

Moreover, in this paper we analyse a collection of optical spectra taken after the former SSS of these and several other SSS-novae returned to quiescence. Several of the spectra we collected are taken in the spectral region in the wavelength range of the H$\beta$ and the He II 4686Å lines and the Bowen blend. The He II 4686Å line requires a ionizing continuum source peaking in the 200–300Å range; it is typical of binaries accreting at high $\dot{m}$, of recent novae and of IPs. The Bowen blend is "pumped" by lines in the UV region. Studying the spectra of planetary nebulae, Bowen (1934, 1935) pointed out that, by a natural coincidence, a few ultraviolet resonance lines of He II, [O III], and N III have almost exactly the same wavelength. Bowen identified *excitation by secondary radiation*, or *line absorption by secondary quanta*: the decays of the above ultraviolet transitions produce the N III emission lines at optical wavelengths at 4640.64, 4641.85 and 4634.13 Å, and a subsequent decay causes two more transitions, N III at 4097.36 and 4103.39 Å. In addition, lines of [O III], C III and more rarely O II are also produced close to 4640-4650 Å, making the so called "Bowen complex" or "Bowen blend" which has by now frequently observed not only in planetary nebulae, but also in X-ray binaries, symbiotic stars and novae. Additional decays, charge-exchange processes and a third "Bowen cascade" in astronomical sources were analyzed by Kastner & Bhatia (1996).

We also followed the development of the nebular, forbidden [O III] lines at 4959 and 5008 Å, in order to monitor the contribution of the nova shell to the optical emission.

In the next section we present a general overview of the objects studied in this paper. Section 3 describes the observation and data analysis, Section 4 focuses on our IP or IP candidate novae, Section 5 describes the optical spectra of the short period novae, Sections 6 focuses on two symbiotic novae, and Section 7 concludes the paper.

**Table 1.** Properties of the novae.

| Object | Outburst maximum | Ampl. ($V$ mag.) | $t_2$ (d) | $t_3$ (d) | $P_{orb}$ (d) | $M_{WD}$ $M_\odot$ | Distance (kpc) | E(B-V) mag. | Class |
|---|---|---|---|---|---|---|---|---|---|
| V4743 Sgr | $2002/09/20^{h1}$ | $\gtrsim 11.2$ | $6^{s}$ | $12^{s}$ | $0.28^{h2}$ | $1.1\text{-}1.2^{h3}$ | $3.90^{+2.37}_{-1.19}$ | $0.12^{h4}$ | CN |
| V2491 Cyg | $2008/04/11^{c1}$ | $\gtrsim 9$ | $4^{s}$ | $16^{s}$ | | $1.3^{c2}$ | $4.53^{+4.81}_{-2.63}$ | $0.43^{c3}$ | CN (RN?) |
| T Pyx | $2011/05/11^{e0}$ | $9.1^{e1}$ | $32^{e1}$ | $62^{e1}$ | $0.076^{e2}$ | $1.25\text{-}1.4^{e3}$ | $2.93^{+0.41}_{-0.33}$ | $0.35^{e2}$ | RN |
| V382 Vel | $1999/05/23^{d1,d2}$ | 14. 3 | $6^{s}$ | $13^{s}$ | $0.1461^{d3}$ | $1.1\text{-}1.2^{d2}$ | $1.71^{+0.18}_{-0.15}$ | $0.2^{d5}$ | CN |
| KT Eri | $2009/11/14^{b1}$ | $\sim 9$ | $\sim 6^{b1,b2}$ | $14.7^{b2}$ | | | $3.69^{+0.53}_{-0.42}$ | $0.08^{b3}$ | CN (RN?) |
| U Sco | 2010/01/28 | $10.1^{e}$ | $1.2^{e}$ | $2.6^{e}$ | $1.23^{g1}$ | $>1.37^{g2,g3}$ | $6.56^{+3.57}_{-1.35}$ | $0.24^{g4}$ | RN |
| HV Cet | $\leqslant$2008/07/17 | $>6$ | | | $\sim 1.7^{a1,a2}$ | | $\sim 2.5^{a1,a3}$ | $0.146^{a4}$ | CN |
| RS Oph | $2006/02/13^{f1}$ | $6.2^{e}$ | $6.8^{e}$ | $14^{e}$ | $453.6^{f2}$ | $\geqslant 1.2^{f4}$ | $2.14^{+0.28}_{-0.22}$, $1.4^{f3}$ | | RN, Symb. |
| V3890 Sgr | 1990/04/27 | $7.4^{e}$ | $6.4^{e}$ | $14.4^{e}$ | $519.7^{e}$ | | $4.36^{+2.36}_{-1.32}$ | | RN, Symb |

Except for HV Cet, the distances have been obtained from the GAIA database using ARI's Gaia Services at http://gaia.ari.uni-heidelberg.de/tap.html. We give also a pre-GAIA alternative estimate for RS Oph, because the symbiotic's nebula and disk can make the parallax determination more uncertain than the nominal uncertainty in the database. The other references are the following: [a1]Beardmore et al. (2012), [a2]Goranskij & Metlova (2009), [a3]Schwarz et al. (2008), [a4]Schwarz et al. (2011), [b1]Hounsell et al. (2010b), [b2]Imamura & Tanabe (2012), [b3]Ragan et al. (2009), [c1]Munari et al. (2011), [c2](Hachisu & Kato 2009), [c3]Rudy et al. (2008), [d1]Gilmore & Kilmartin (1999), [d2]Della Valle et al. (2002), [d3]Bos et al. (2001), [d4]Shore et al. (2003), [d5]Cardelli et al. (1989) and Augusto & Diaz (2003) for discussion, [e0]see for a review Schaefer et al. (2013), [e1]from Fig. 4 of Surina et al. (2014), [e2]Patterson et al. (1998), [f1]Hounsell et al. (2010a), [f2]Brandi et al. (2009), [f3]Barry et al. (2008), [f4]Nelson2008, [g1]Schaefer (1990),[g2]Thoroughgood et al. (2001),[g3]Hachisu et al. (2000),[g4]Burstein & Heiles (1982), [h1]Haseda et al. (2002), [h2]Kang et al. (2006); Richards et al. (2005); Wagner et al. (2005), [h3]Nielbock & Schmidtobreick (2003), [h4]Zemko et al. (2016), [s]Strope et al. (2010).

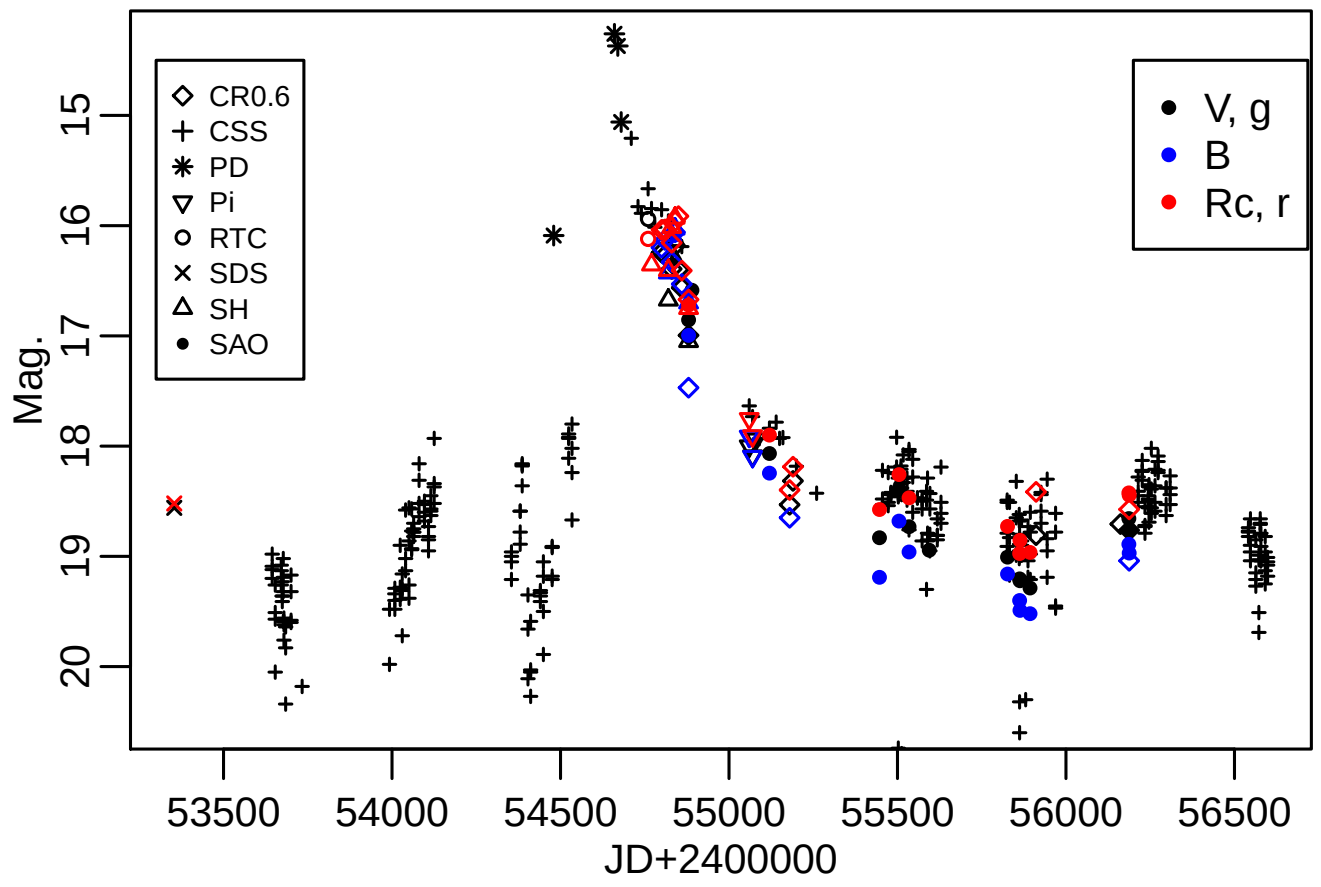

**Figure 1.** The long-term light curves of HV Cet in the $B$ (blue), $V$, $g$ (black) and $Rc$, $r$ (red) filters, covering the nova outburst. Since the nova was intensively monitored during the decline, in the interval between JD 2454600 and JD 2455300 we plotted the 10 days-averaged magnitudes. **CR0.6** — Goranskij & Metlova (2009), **CSS** — the Catalina Sky Survey (CSS) data (Drake et al. 2009a), **PD** — Prieto et al. (2008), **SDS, RTC** — $g$, $r$ magnitudes from SDSS and RETROCAM from Prieto et al. (2008), **SAO** and **Pi** — $BVRc$ photometry performed by V. Goranskij with the 1-meter Zeiss telescope of the Special Astrophysical Observatory of Russian Academy of Sciences (SAO RAS) and with the Sternberg Astronomical Institute (SAI) Crimean Station 0.5-m Maksutov meniscus telescope (AZT-5). **SH** - $BVRc$ photometry performed by S. Shugarov with the 0.5-m telescope G1 of the Stara Lesna Observatory, Slovakia.

## 2 A BRIEF HISTORY OF OUR TARGETS

In Table 1 we list the physical parameters known for each nova in our program. The first are the IP candidates, then we listed the other novae in order of growing orbital period. We have observed one recurrent nova below the period gap, two classical novae with periods of few hours, two novae (recurrent and not) with 1-2 days' periods and two more suspected to have orbital periods in the same range, and finally two recurrent symbiotic novae.

### 2.1 V4743 Sgr

V4743 Sgr, a moderately fast nova, was discovered in outburst on 2002 September 20 by Haseda et al. (2002) when it was close to the 5th magnitude. V4743 Sgr continued nuclear-burning for at least for 1.5 years after the outburst (Rauch et al. 2010). A modulation of the X-ray light curve on the timescale of 22 min, reported by several authors (see Paper II, and Ness et al. 2003; Leibowitz et al. 2006; Dobrotka & Ness 2010), revealed that this binary system hosts a magnetic WD. The orbital period (~6.7 h) and a proposed beat period between the orbital and the spin one (~24 min) were detected in the optical band by Kang et al. (2006), Richards et al. (2005) and Wagner et al. (2005).

The distance to the nova is uncertain. The estimates range from 1.2 kpc (Nielbock & Schmidtobreick 2003) to 3.9 kpc (Vanlandingham et al. 2007). Nielbock & Schmidtobreick (2003) found that the distance cannot exceed 6 kpc.

### 2.2 V2491 Cyg

V2491 Cyg (Nova Cyg 2008b) was discovered by Hiroshi Kaneda on 2008 April 10.728 at $V$=7.7 (Nakano et al. 2008). The nova was very fast and its light curve showed an unusual feature – a short "re-brightening" at the end of April. Jurdana-Sepic & Munari (2008) discovered the optical counterpart of the nova at pre-outburst magnitude V≃17.1, with a short dimming before the outburst. Although no previous outburst was detected, V2491 Cyg share many properties with RNe (see discussion in Paper I). Pre-outburst X-ray observations of V2491 Cyg revealed that the WD was accreting at a high rate (Ibarra et al. 2009) and the high WD atmospheric temperature measured during the SSS phase indicates that the WD is also very massive (Ness et al. 2011). There are also several indications that the WD in V2491 Cyg is magnetic (see Paper I, and Hachisu & Kato 2009; Ibarra et al. 2009; Takei et al. 2011). The best (albeit indirect) proof of an IP is a modulation of the soft X-rays with the rotation period of the WD. This period may be the ~38 min one measured both in the optical and in X-rays, but it was unclear whether it was stable (in addition to Paper I, see Shugarov et al. 2010; Darnley et al. 2011; Ness et al. 2011). However, in more recent, still unpublished *XMM-Newton* data this period has been detected again with higher confidence and it does seem to be indeed stable. Thus, it is very likely to be the rotation period of the WD (Orio et al. 2018, in preparation). The orbital period of the system is not known. Baklanov et al. (2008) detected a modulation with a period of 0.09580(5) days between 10 and 20 days after the outburst, that has not been measured again in more recent works (Shugarov et al. 2010; Page et al. 2010; Darnley et al. 2011).

### 2.3 T Pyx

T Pyx is a RN that had outbursts in 1890, 1902, 1920, 1944, 1967 and 2011 (see Gilmozzi & Selvelli 2007; Selvelli et al. 2008; Schaefer 2010; Schaefer et al. 2013, for a review). The explosions are typically slow, with $t_2$=32 and $t_3$=62 days, respectively (Schaefer 2010). The outburst light curves show day-to-day oscillations of 0.5–1.0 mag. Patterson et al. (1998) has shown that apart from a dominant stable periodicity of 0.0762264(4) d, which has been attributed to orbital variation, there are weaker signals at 0.1098 and 1.24 d. The orbital period increases on a timescale $\frac{P}{\dot{P}} = 3 \times 10^5$ yr (Patterson et al. 2014) and from the orbital period change Patterson et al. (2014) inferred that, although the WD in T Pyx is very massive (~1.35 $M_\odot$ Selvelli et al. 2008), the ejected mass exceeds the amount of accreted material. The same result was found by Selvelli et al. (2008) by modelling IUE observations. The X-ray spectra show that the central source never became completely visible, probably because of obscuration of the ejecta (Tofflemire et al. 2013).

### 2.4 V382 Vel

The outburst of V382 Vel was discovered independently by P. Williams and A. C. Gilmore on 1999 May 22 (Williams et al. 1999). It was one of the brightest nova ever observed, with $V_{\rm max} \sim 2.3 \pm 0.1$ (Della Valle et al. 2002). V382 Vel was classified as a fast ONe nova (Woodward et al. 1999). The high optical luminosity at maximum and the short turn-off time of the super-soft X-rays (Ness et al. 2005) implies that the nova hosts a massive WD, likely in the mass range 1.1–1.2 $M_\odot$ (Della Valle et al. 2002). Della Valle et al. (2002) estimated the distance to be 1.70±0.34 kpc, while Shore et al. (2003) found a value of 2.5 kpc. Bos et al. (2001) reported that the orbital period of the binary is 0.14615(1) days, which was later confirmed by Balman et al. (2006).

### 2.5 KT Eri

KT Eri was discovered by K. Itagaki on 2009 November 25.536 UT (Itagaki 2009) at 8.1 mag after maximum-light. From the Catalina Sky Survey (CSS) data Drake et al. (2009c) found that the outburst occurred between 2009 November 10.41 UT and 18 UT and Hounsell et al. (2010b) constrained the time of the outburst to 2009 November 14.67±0.04 UT using the USAF/NASA Solar Mass Ejection Imager (SMEI) on board the Coriolis satellite. The nova was very fast: $t_2$ and $t_3$ were ~6 days (Hounsell et al. 2010b; Imamura & Tanabe 2012) and 14.7 days (Imamura & Tanabe 2012), respectively. Jurdana-Šepić et al. (2012) studied historical light curves of KT Eri and found that the progenitor system had $\langle B \rangle = 14.7 \pm 0.4$ and most probably contained an evolved secondary. Although several properties of KT Eri make it similar to RNe, no previous outburst was detected (for a review see Jurdana-Šepić et al. 2012). The authors also found two periodicities of 737 and 367 days. In the Swift archive we find that KT Eri was a very luminous SSS until about 300 days after the outburst (see also Ness et al. 2014).

### 2.6 U Sco

U Sco is a well studied recurrent nova (for a review see Schaefer 2010; González-Riestra 2015). Ten outbursts of U Sco were detected: in 1863 and then every 10±2 years, with the last one on 2010 January 28 (Schaefer et al. 2010). The orbital period of the system is 1.2344 days (Schaefer 1990). Nova eruptions in U Sco occur on very short timescales: rise to maximum in 6-12 hours, $t_2$ and $t_3$ are 1.2 and 2.6 days, respectively and return to quiescence in 67 days (Schaefer 2010; Pagnotta et al. 2015). U Sco hosts a very massive WD (>1.37 $M_\odot$ Hachisu et al. 2000; Thoroughgood et al. 2001) and the ejecta velocity reaches 10 000 km $s^{-1}$ (González-Riestra 2015). At maximum it is usually around $V$=7.5 mag and in quiescence ~ 17.5 mag and 18.9 during the total eclipse (Schaefer 2010). The short SSS phase was observed in the last outburst only through Thomson scattering reflected emission, because the system has high inclination, and described by Ness et al. (2012) and by Orio et al. (2013).

### 2.7 HV Cet

CSS081007:030559+054715 (HV Cet) was discovered by the Catalina Real-time Transient Survey (CRTS) on 2008 October 07 (Drake et al. 2009b). Observed as it was coming out of the Sun, it had brightened by about 4 magnitudes in less than one year and it was classified as a possible RN of the ONe class, however the time of the maximum was unknown (Pejcha et al. 2008). We collected photometric observations from several surveys and observers in order to bridge the observational gap as much as possible and present a previously unpublished outburst light curve of HV Cet in Fig. 1, including still unpublished photometric measurements. Unfortunately, however, the nova was behind the Sun during the peak and the initial decline, possibly even while it declined by several magnitudes. There is evidence that it may have been a fast nova, because of the high ejecta velocity.

Several possible periodical variations in X-rays, UV and optical were reported during the late outburst phase in which the nova was monitored. Beardmore et al. (2012) investigated the X-ray, UV and optical evolution including the American Association of Variable Star Observers (AAVSO) light curve and GALEX observations and reported a modulation with a period of 1.77 days. Goranskij & Metlova (2009) monitored the nova at quiescence in $B, V, Rc$ filters and measured two periods: 1.694(5) and 0.6106(6) days, which surprisingly were not detected in pre-outburst light curves (Drake et al. 2009b). Drake et al. (2009b) also measured optical polarization, and no significant circular polarization was found.

### 2.8 RS Oph

RS Ophiuchi is a symbiotic RN that undergoes thermonuclear runaways every 9 to 21 years. Six outbursts were detected (in 1898, and no 1933, 1958, 1967, 1985 and 2006) and another two, in 1907 and 1945 and were missed due to a seasonal gap, but were later partially detected by Schaefer (2010) and Adamakis et al. (2011). RS Oph hosts a M0–2 III giant component of mass 0.68–0.80 $M_\odot$ (Brandi et al. 2009) and a short recurrence time of the outbursts implies a massive WD with mass of at least 1.2 $M_\odot$ (Mikołajewska & Shara 2017; Brandi et al. 2009). The orbital period of the system is 453.6 days (Brandi et al. 2009). High mass transfer rate (Booth et al. 2016) and evidence of a massive WD make it a good candidate type Ia SN progenitor. During the most recent outburst RS Oph was intensively observed from radio to X-rays. The X-ray evolution from a hard source originating in the shocked circumstellar material to an SSS were described by Ness et al. (2007), Nelson et al. (2008) and Ness et al. (2009). It was imaged with the *Hubble Space Telescope* (Ribeiro et al. 2009), and observed in infrared by Chesneau et al. (2007). Radio interferometry (O'Brien et al. 2006) revealed bipolar outflows interacting with a dense circumstellar material (see e.g. Bode et al. 2006; Sokoloski et al. 2008; Eyres et al. 2009; Vaytet et al. 2011, and references therein). Iijima (2009) followed the spectral evolution of RS Oph after the 2006 outburst and found a short-lived flare-up of He I emission lines, which they attributed to a helium flash (however, we note, without dwelling into the subject, that this interpretation is very arguable). Many emission lines in the post-outburst spectra had a central narrow component, blue- and red-shifted wings, and a very broad component (Iijima 2009).

The quiescence optical spectra of RS Oph are characterized by a complex structure of the Balmer emission lines and a pronounced variability on different time-scales. The H$\alpha$ emission line has a narrow component (the full width at half maximum, or FWHM is ~220 km $s^{-1}$), a highly variable broad one (FWHM>2000 km $s^{-1}$), and a narrow absorption component (Zamanov 2011), slightly blue-shifted with respect to the narrow one, which disappears during the outbursts (Brandi et al. 2009). This absorption component does not follow the motion of the stellar components of the binary system (Brandi et al. 2009). The H$\alpha$ line also has a very broad base with full width at zero intensity, or FWZI, ≈4600 km s$^{-1}$ (Zamanov 2011). Worters & Rushton (2014) performed high time resolution spectroscopy of RS Oph and found that the equivalent width (EW) of the double-peaked H$\alpha$ is variable on the timescale as short as 2 min, while the ratio of the blue to red peak remains approximately the same. The authors did not find a correlation between the variability of the H$\alpha$ line and the strength of other lines, proposing that the variation is intrinsic to H$\alpha$ and is probably related to changes in the photoionization of the nebula. To explain the variability of the broad component of H$\alpha$ Zamanov (2011) proposed several mechanisms, including blobs of matter, variable disk winds and an asymmetric Keplerian disk. Pavlenko et al. (2016) also proposed accretion disc variability as the explanation of the variation of the spectral energy distribution and intensity of the emission lines on ≃1 day timescales. Even if it is not clear whether the secondary does fills its Roche lobe, RS Oph almost certainly hosts an accretion disk, like probably many other symbiotics (see references and discussion in Orio et al. 2017).

**Table 2.** Summary of observations.

| Object | Date of obs. | $\Delta T^a$ (yr) | Exposure time (s) | Instrument$^b$ |
|---|---|---|---|---|
| | | **Photometry** | | |
| V4743 Sgr | 2015/10/04-2015/12/26 | 13.1 | 18000 | Kepler |
| V2491 Cyg | 2015/06/16-18<br>2015/07/17-23 | 7.2<br>7.3 | 120-180<br>90-300 | WIYN 0.9m, *UBVRI*<br>SAI-Crimea 1.25m , *C* |
| | | **Spectroscopy** | | |
| V2491 Cyg | 2008/06/08 23:59:10–00:53:04<br>2009/07/30 22:38:55–23:05:31<br>2012/08/17 00:03:40–00:20:13<br>2015/08/23 18:31:25–20:17:16 | 0.2<br>1.3<br>4.4<br>7.4 | 600,1200,2400<br>420,440,600<br>2×900<br>16×300 | Zeiss 1000 UAGS 325/4<br>BTA SCORPIO VPHG1200G<br>BTA SCORPIO VPHG550G<br>BTA SCORPIO VPHG1200B |
| T Pyx | 2012/12/06 23:40:57<br>2014/03/01 | 1.6<br>2.8 | 620<br>480 | SALT RSS PC00000 PG2300<br>SALT HRS R=14,000 |
| V382 Vel | 2012/07/10 17:15:40<br>2013/04/30 21:08:38–21:58:57 | 13.1<br>13.9 | 1700<br>4×990 | SALT RSS PC00000 PG2300<br>SALT RSS PC00000 PG2300 |
| KT Eri | 2010/01/22 17:51:47-18:52:46<br>2010/12/06 21:05:53-21:09:56<br>2011/12/23 20:13:59<br>2012/10/19 00:31:50-00:39:25<br>2012/11/22 20:52:48 | 0.2<br>1<br>2<br>3<br>3 | 4×1200<br>150,462<br>695<br>2×400<br>1000.2 | Zeiss-1000 UAGS 325/4<br>BTA SCORPIO VPHG1200G<br>BTA SCORPIO VPHG550G<br>BTA SCORPIO VPHG550G<br>SALT RSS PC00000 PG2300 |
| U Sco | 2013/05/01 02:37:52–02:58:03<br>2013/05/01 21:43:02–22:01:39<br>2013/05/28 00:45:19–01:03:58<br>2013/06/10 23:54:14–00:12:53 | 3.2<br>3.2<br>3.3<br>3.4 | 2×1194<br>2×1100<br>2×1100<br>2×1100 | SALT RSS PC00000 PG2300 PL0150N001<br>SALT RSS PC03850 PG0900<br>SALT RSS PC03850 PG0900<br>SALT RSS PC03850 PG0900 |
| HV Cet | 2013/11/05 | 3.3 | 2×1060 | SALT RSS PC00000 |
| RS Oph | 2011/09/15 18:18:31–19:27:44<br>2011/09/16 19:57:16–21:27:48<br>2011/09/19 19:48:33–20:52:49<br>2011/09/20 20:07:59–20:52:11<br>2012/09/14 19:03:45–19:45:44<br>2012/09/16 18:39:09–19:23:03 | 5.6<br>5.6<br>5.6<br>5.6<br>6.6<br>6.6 | 600,1200,1800<br>4×1200<br>2×600,2×1200<br>4×600<br>2×1200<br>1200,2×600 | SAAO grating 4<br>SAAO grating 4<br>SAAO grating 5<br>SAAO grating 5<br>SAAO grating 4<br>SAAO grating 5 |
| V3890 Sgr | 2011/09/16 18:14:56–19:43:29<br>2011/09/17 18:33:16–20:40:50<br>2011/09/19 18:14:35–19:38:19<br>2011/09/20 18:00:58–19:33:30<br>2012/09/15 20:18:08–21:20:32<br>2012/09/18 20:21:46–21:48:19 | 21.4<br>21.4<br>21.4<br>21.4<br>22.4<br>22.4 | 1200,2×1800<br>6×1200<br>4×1200<br>4×1200<br>2×1800<br>3×1800 | SAAO grating 4<br>SAAO grating 5<br>SAAO grating 5<br>SAAO grating 5<br>SAAO grating 4<br>SAAO grating 5 |

$^a$Time since the outburst maximum, $^b$RSS — Robert Stobie Spectrograph, HRS — High-Resolution Spectrograph, PC00000 — clear filter, PC03200, PC03400, PC03850 — UV blocking filters, PC04600 — blue blocking filter.
Gratings wavelength ranges: PG0900, 3200–7100Å; PG2300, 4200-5600 Å; VPHG1200G, 4000-5800Å; VPHG550G, 3700-7800Å; VPHG1200B, 3700-5500Å; SAAO grating 4, 4000–4900Å; SAAO grating 5, 6200–7000Å.

### 2.9 V3890 Sgr

The symbiotic RN V3890 Sgr was discovered as a CN in 1962 (Duerbeck 1987) and on 27 April 1990 it underwent a second outburst (Kilmartin et al. 1990). Although it did not initially belong to our program because it was sparsely observed optically and there is only one serendipitous X-ray observation at a late outburst phase (Orio et al. 2001b), we obtained a low resolution spectrum that we include for comparison with RS Oph.

## 3 OBSERVATIONS AND DATA ANALYSIS

### 3.1 The photometry

We proposed Kepler observations of V4743 Sgr in the framework of the K2 campaign (field #7). The details of the observations are in Table 2. Both short (2 min) and long (30 min) cadence light curves were obtained. The analysis of the short-cadence light curve is now in progress and we will discuss here only the long-cadence light curve.

Photometric observations of V2491 Cyg were performed with the 0.9-m WIYN telescope of the Kitt Peak National Observatory (KPNO) and the 1.25-m telescope of the Crimean Station of the Sternberg Astronomical Institute of Moscow University. During the KPNO observations we measured the mean magnitudes of V2491 Cyg in the U, B, V, R, and I filters and monitored the nova in the V filter. For better accuracy, the SAI Crimean Startion observations were performed without filter. For absolute magnitude calibration we used the 112 822 star from the Landolt Equatorial Standards catalog (Landolt 1992). The mean absolute magnitudes and the errors are in Table 3.

### 3.2 The spectra

Spectroscopic observations were performed with the 10-meter Southern African Large Telescope (SALT), with the Cassegrain Spectrograph on the 1.9-m telescope of the South African Astronomical Observatory (SAAO)[1], with the 6-meter Big Azimuthal Telescope (BTA) and with the 1-meter Zeiss telescope of the Special Astronomical Observatory of the Russian Academy of Sciences (SAO RAS). The observations performed with the SAAO 1.9-m telescope will simply be referred to throughout the paper as the SAAO observations. The observation dates, exposure times and specific instruments used are summarized in Table 2.

Two different SALT instruments were used: the High-Resolution Spectrograph (HRS) and the Robert Stobie Spectrograph (RSS). The HRS was used in its low resolution mode (resolving power $R = 14,000$). All the SALT observations were performed in long-slit mode; some of them were obtained with the PG2300 grating to cover the 4200-5600 Å range where our main lines of interest are, while others were taken with the PG900 grating to cover the 4200-7100 Å range, including the prominent H$\alpha$ line. The width of the slit was $1''$ in both the SALT and the BTA observations, $1.51''$ in the SAAO ones. The SALT and the BTA spectra were flux calibrated with spectrophotometric standard stars, although SALT is a telescope with variable pupil, which makes the absolute flux calibration impossible. For the SALT spectra, therefore, we give only relative fluxes. All the flux-calibrated spectra were also dereddened with the values of E(B-V) given in Table 1. The 1.9 m SAAO spectra were not flux calibrated because spectrophotometric standards were not observed in those nights. Nevertheless, by making use of relative counts, we were able to compare the EWs of the lines and the flux ratio of lines next to each other. The emission line fluxes and EWs were calculated by means of fitting Gaussian components (when the shape of a line was complex we used several Gaussians). For the Bowen blend, we followed McClintock et al. (1975), as illustrated in Fig. 5, using the same main component lines, including several lines of O II at about 4687 Å that these authors introduced for X-ray binaries in which gaseous matter is irradiated by X-rays (at higher temperature than the SSS in novae, at almost 10 $R_\odot$ distance).

In order to calculate the errors of the flux and EW of each line we repeated the following procedure 1000 times: we introduced a noise normally distributed with the same standard deviation as in the continuum to the Gaussian line profile and repeated the Gaussian fitting. The values we give for the flux and EW are the mean of the set of these 1000 measurements and the errors were found from its standard deviation.

## 4 EXPLORING THE MAGNETIC NATURE OF V4743 SGR AND V2491 CYG

### 4.1 V4743 Sgr

The 30 min cadence light curve of V4743 Sgr was extracted using the K2SFF tool (Vanderburg & Johnson 2014)[2]. The top-left panel of Fig. 2 shows this light curve, which is highly variable on timescale of several days, with an amplitude exceeding two magnitudes, but without a definite periodicity. We divided the light curve in 1 day segments and removed the long-term variations with a third-order polynomial fitting for each segment. The de-trended light curve is presented in the top-right panel of Fig. 2. We then applied the Lomb-Scargle algorithm (Scargle 1982) to search periodic modulations. The middle panels show spikes in the LSPs, which correspond to the orbital period (6.684 h) and to the beat-period between the orbital and the spin one (23.65 min). Taking into account these two periodicities and the period found in the X-ray light curves (Paper II, and Dobrotka & Ness 2010), the relation $1/P_{\rm spin} - 1/P_{\rm orb} = 1/P_{\rm beat}$ is perfectly satisfied, thus confirming the result of Kang et al. (2006), who measured the beat of the orbital and WD rotation periods. This implies the stability of the WD spin period and brings even stronger evidence for our classification of V4743 Sgr as an IP in Paper II.

[1] http://www.saao.ac.za/wp-content/uploads/sites/5/spmanv6.5.pdf

[2] https://archive.stsci.edu/k2/hlsp/k2sff/search.php

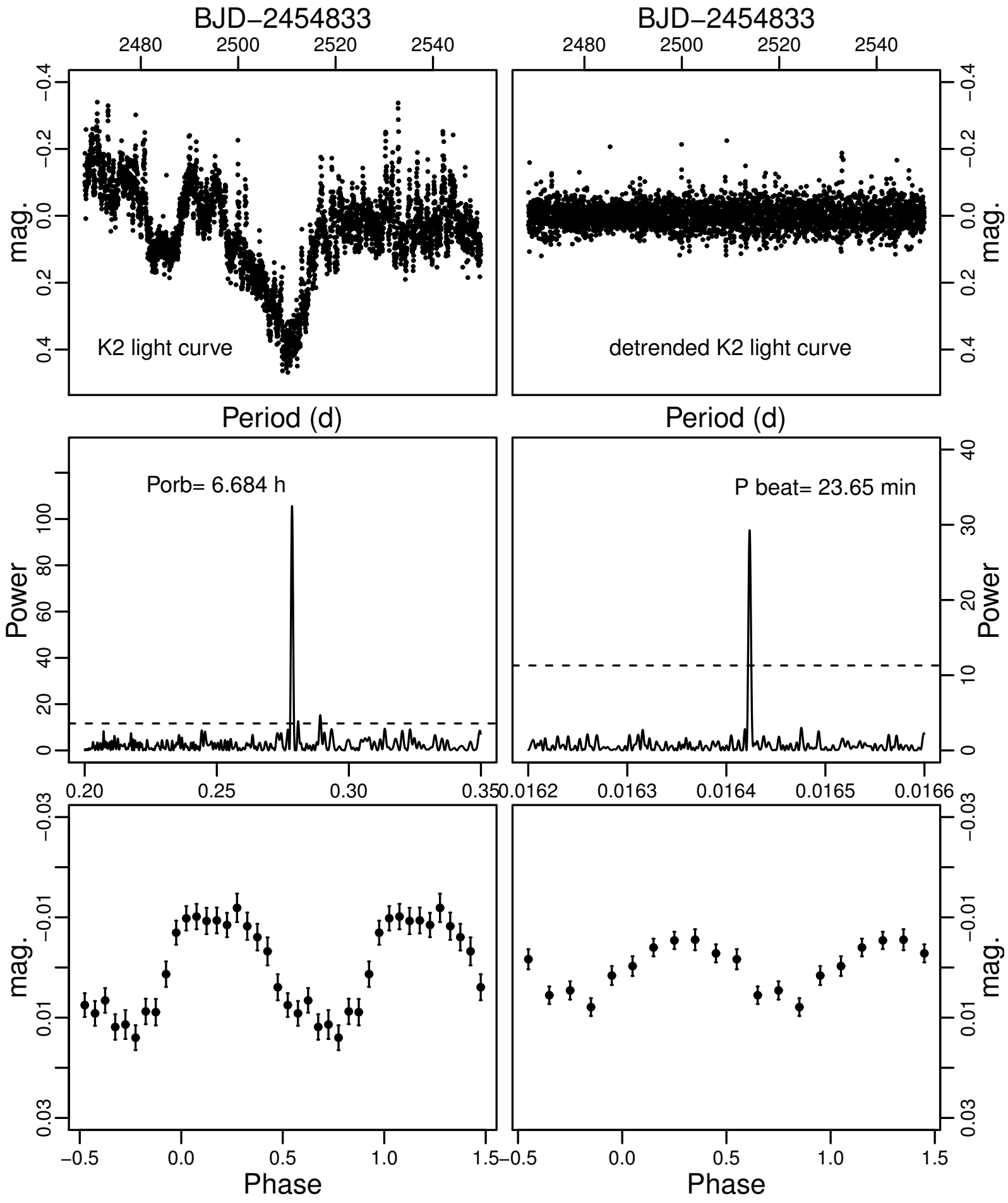

**Figure 2.** Top: the original (left) and de-trended Kepler light curves of V4743 Sgr. Middle: LSP of the detrended Kepler light curve showing the orbital (left) and the beat (right) period. The horizontal dashed line shows the 0.3% false alarm probability level. Bottom: the light curve folded with the orbital (left) and the beat (right) periods.

### 4.2 V2491 Cyg

Our X-ray observations of V2491 Cyg (Paper I) revealed that it is likely to be an IP (see Section 2.2). We combined the KPNO and SAI-Crimean photometric observations, removed possible long-term variations from each nightly light curve, using 1–4 orders polynomial fitting (the order of polynomial function was chosen depending on the length of observation at each night) and performed a timing analysis with the Lomb-Scargle algorithm. We also analyzed the combined data set with the Deeming (1975) method of descrete Fourier transform for arbitrarily distributed time series. In this analysis, the long-term variations were decomposed into several harmonic components, which were removed from the series using a prewhitening procedure. The peaks around two frequencies, respectively 0.0276 d and 0.032 d, appear independently on the method we apply. We found out that the frequencies close to the one with the highest probability in each group are daily aliases. We did not find any reliable modulation on longer timescales, because on periods longer than 0.045 d different methods yield slightly different results, and we concluded that with the nature and quality of the data at hand, we cannot detect such frequencies with confidence. The data also do not allow us to investigate the peaks around 0.032 d. These peaks are due to four different frequencies that are most probably different sinusoidal components of a weak, quasi-periodic signal.

The highest peak is the one at 0.0276 d (36.24 minutes), and although the signal is low, this finding may be very important in the context of the magnetic interpretation of V2491 Cyg presented in Paper I. In fact, this period is only slightly shorter than the 0.0266 d (38 minutes) one detected in the X-rays, thus a possible interpretation is that it is the beat of the rotational and orbital period, as a result of reprocession of the X-rays from the surface of the secondary on the magnetic WD. If this interpretation is correct, taking our measurements of the putative spin and beat periods into account, the orbital period should be about 0.7 days (16.8 hours), which is too long to be detectable in our observations.

The optical spectra of V2491 Cyg are presented in Figures 4, 5 6. All the spectra were flux calibrated and dereddened. Figure 4 shows the evolution of the optical spectrum from the outburst to quiescence. Since the outburst spectra of V2491 Cyg were studied in detail by Ribeiro et al. (2011), we focus here on the spectra obtained after one year from the explosion. We only note here that we confirm that the emission lines in outburst have a complex saddle-like struc-ture with wings, extended up to 4600 km s$^{-1}$, and following the discussion of Ribeiro et al. (2011) our outburst spectrum is consistent with their model A (polar blobs with an equatorial ring), but it has a stronger contribution from the component associated with the equatorial ring.

**Table 3.** *U,B,V,R,I* magnitudes of V2491 Cyg.

| Filter | *U* | *B* | *V* | *R* | *I* |
|---|---|---|---|---|---|
| Mag. | 18.27±0.07 | 18.72±0.08 | 18.03±0.03 | 17.72±0.03 | 17.41±0.02 |

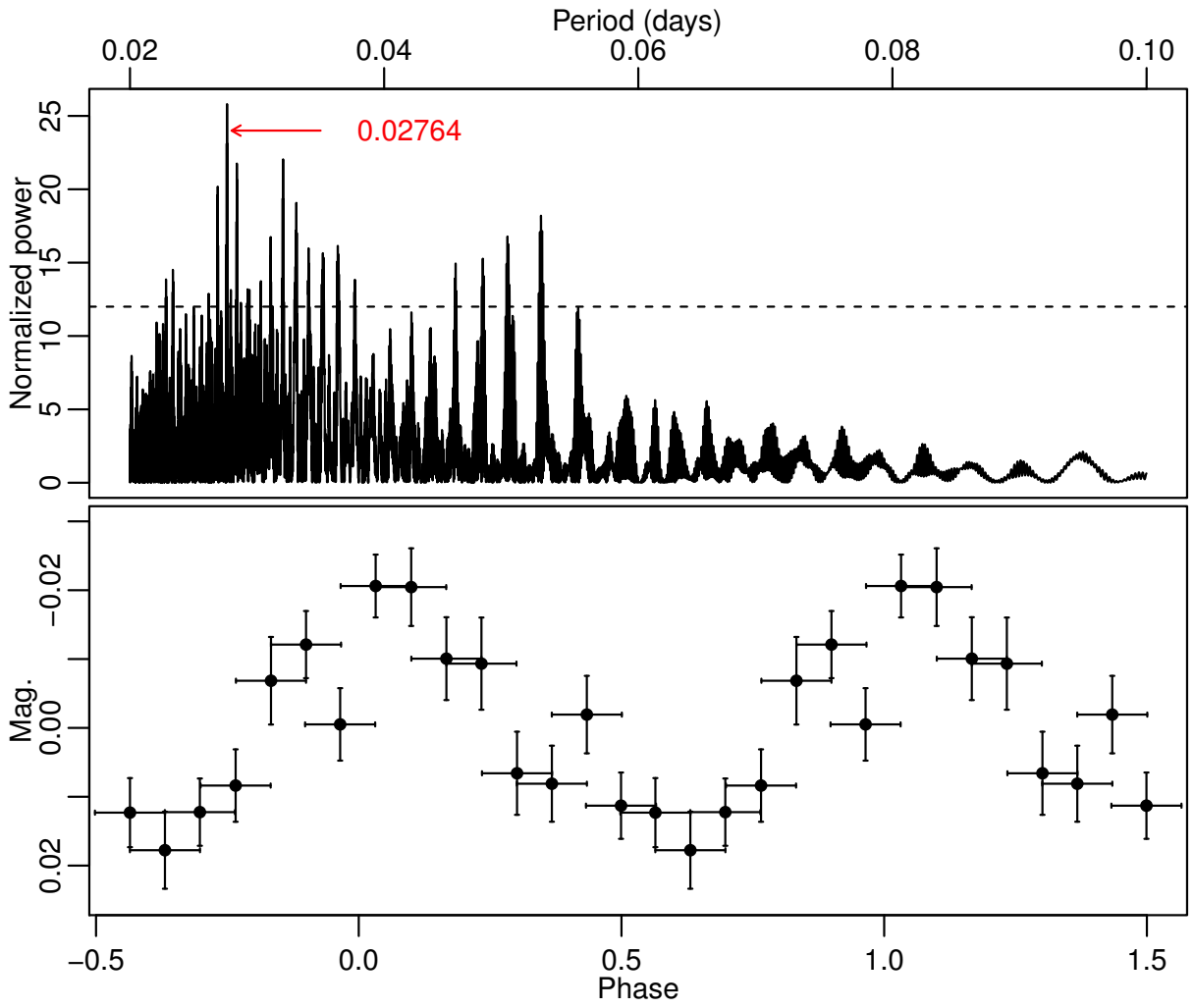

**Figure 3.** Top: LSP of the KPNO and SAI Crimean Station observations of V2491 Cyg. The highest peak marked with the red arrow corresponds to the 0.02764 days period. The horizontal dashed line represents the 0.3% false alarm probability level. Bottom: the light curve folded with the 0.02764 days period.

The [O III] line, originating in the nova shell, was weak but still visible in the spectra obtained 4.3 and 7.3 years after the outburst. The main features of the quiescent spectra are the Balmer lines, the He II λ4686 line and the Bowen blend λλ4640 – 4650. All the emission lines in the quiescent spectra are single-peaked (except the [O III] line, originating in the nova shell), which is most probably due to low orbital inclination.

The fact that the orbital period has never been detected in this system, neither photometrically nor spectroscopically, may imply that the inclination is lower than than ∼ 60 degrees. Limits on the inclination can also be derived from the lack of a double peaked profile. The resolution of the optical spectra, presented in this work, was 5.6 Å in 2009, 15.8 Å in 2012 and 6 Å in 2015. Such resolution would allow to measure line separation of 190 – 360 km s$^{-1}$, however we did not detect any line splitting due to the Keplerian motion in the accretion disk. If V2491 Cyg is an IP with $M_{WD}$=1.2 $M_\odot$, the Keplerian velocity at the synchronization radius corresponding to the proposed 38 min rotation period is 760 km s$^{-1}$. In case of non-magnetic accretion, the maximum plasma velocity in the accretion disk is even higher. If the Balmer lines in the V2491 Cyg spectra originate from the accretion disk, the absence of the double-peaked profiles gives an upper limit for the inclination of arcsin(190/760) = 14 degrees.

In order to analyse the evolution and the shape of the emission lines of V2491 Cyg we localized the regions of the He II λ4686 and the Balmer lines and attempted to fit them with a number of Gaussians after removing the continuum using the `continuum` task in `IRAF`. The result of the fit is marked with the red solid line and the Gaussian components are marked with the black dashed lines in Fig. 5 and 6.

An interesting fact that we want to underline is that almost all the Balmer lines and the He II λ4686 line in the 2012 and 2015 spectra have two components, red and blue-shifted, with the same velocity ∼760 km s$^{-1}$ (see Fig. 5, 5). Double or multiple components in the emission lines are quite typical of IPs (see for instance Ferrario 1991), and in fact we notice that this spectrum resembles that of GK Per, a well-known IP, where similar emission line components were detected during the dwarf nova outburst and were interpreted as emission from the accretion curtain near the magnetized WD (Morales-Rueda et al. 1999; Bianchini et al. 2003). Although some of these line components in V2491 Cyg are weak and are comparable with the noise level, the same velocities and their presence in two spectra, obtained 3 years apart, indicate that they must be real. This analogy with GK Per gives further indication in favor of the IP model for this nova. One may argue that, if the nova shell was larger than the slit width, we may have lost light with zero radial velocity and detected only the light from the material moving from and towards the observer. The slit width was 1" and assuming the same expansion velocity measured during the outburst, 4600 km s$^{-1}$, and a distance of 10 kpc, the upper limit for the shell angular dimension is only 0.38", so we can exclude this possibility. The line components can also be the effect of visibility of the accretion disk at different orbital phases, but since our exposures were most likely shorter than the orbital period, since we did not detect any reliable modulation with periods in the range 0.05–0.2 d, this explanation cannot be accepted.

Another possible explanation for the two components in the emission lines, other than the IP nature, is the presence of a wind from the binary system, however we would expect a broad line component, rather than two narrow components with the same velocity. A third possibility, that we cannot rule out at this stage, is that the two components are due to a bipolar outflow.

## 5 THE SPECTRA OF THE NON-SYMBIOTIC NOVAE

### 5.1 T Pyx

Figure 7 shows the spectrum of T Pyx obtained on 2014 March 3 with the SALT high resolution spectrograph (HRS) and the comparison with the RSS spectrum obtained on 2012 December 12 by Tofflemire et al. (2013), in the 4900-5050 Å region (top panel). We show also a proposed multi-component fit for the strongest lines, He II $\lambda$4686 and H $\beta$ in the middle panel, [O III] in the bottom panel. In this nova, the nebular lines of [O III] appear to be much stronger than in other novae observed at comparable post-outburst epochs, as the next spectra we show indicate. In the insets we zoomed into the H$\beta$ and He II lines. All the lines have a complicate structure with a broad component at zero velocity and some narrow components shifted by ±200 – 1200km s$^{-1}$.

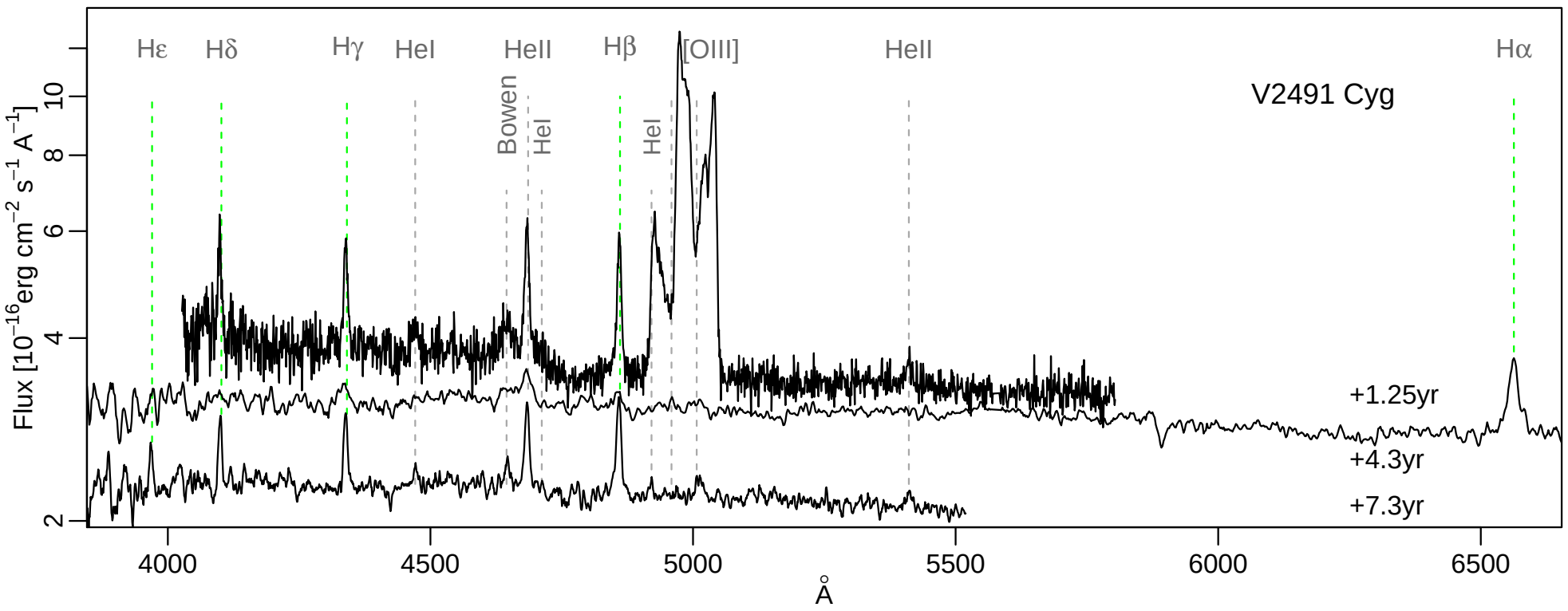

**Figure 4.** The evolution of the optical spectra of V2491 Cyg from outburst to quiescence. Time elapsed since the outbursts are marked near each spectrum. The spectrum obtained 4.3 years after the outburst was shifted upwards by 30% for visibility.

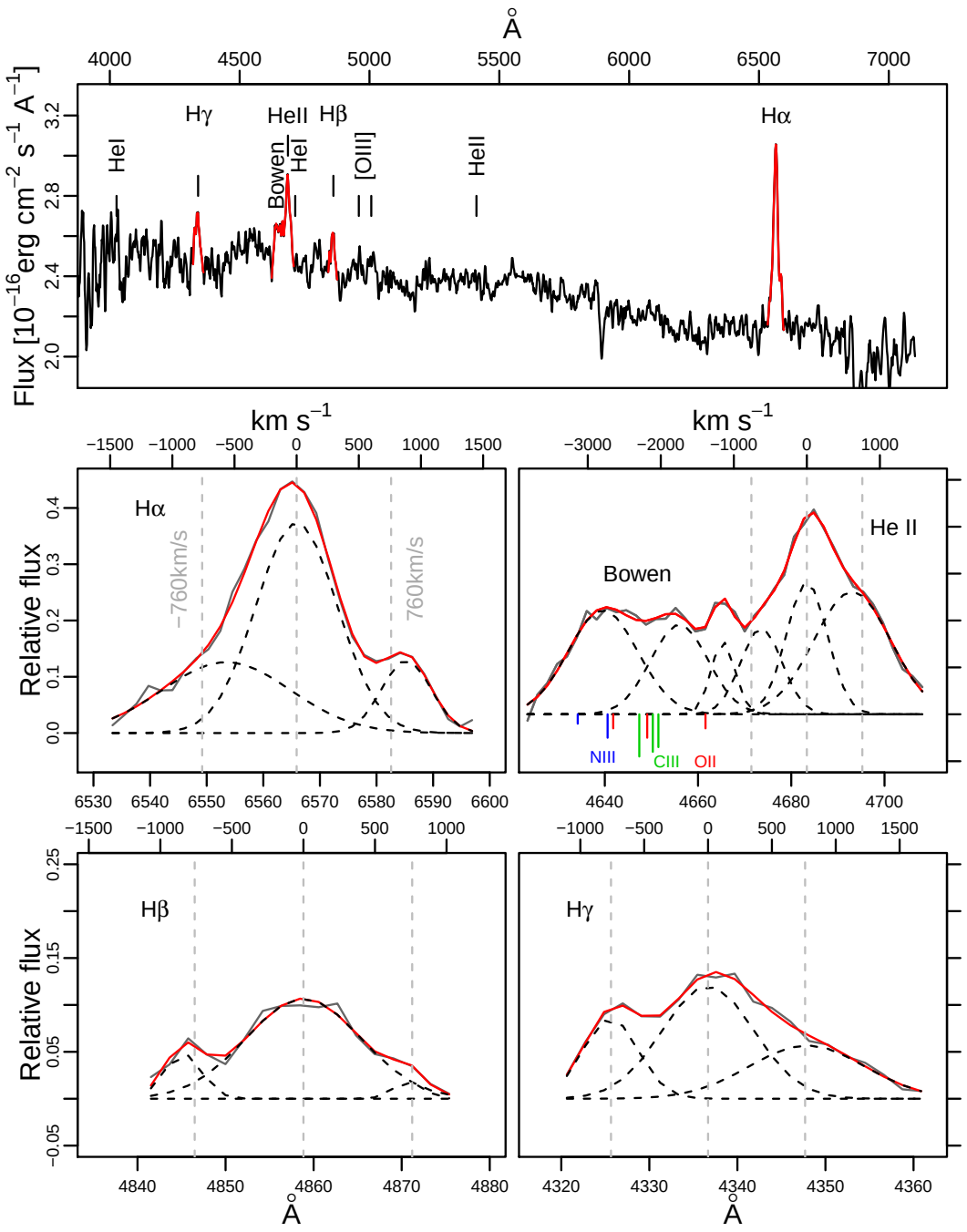

**Figure 5.** Top: The flux calibrated spectrum of V2491 Cyg, obtained in 2012. The strongest emission lines are marked. The result of the fit of the emission lines with Gaussians component is plotted in red. Bottom: the regions of the Balmer lines, the Bowen blend and the He II line after the subtraction of the continuum. The red lines show the result of the fit and the black-dashed lines show the Gaussian components that were introduced, for the main components of the Bowen blend we followed McClintock et al. (1975). The rest wavelength of the components of the fit for lines that constitute the Bowen blend is labeled.

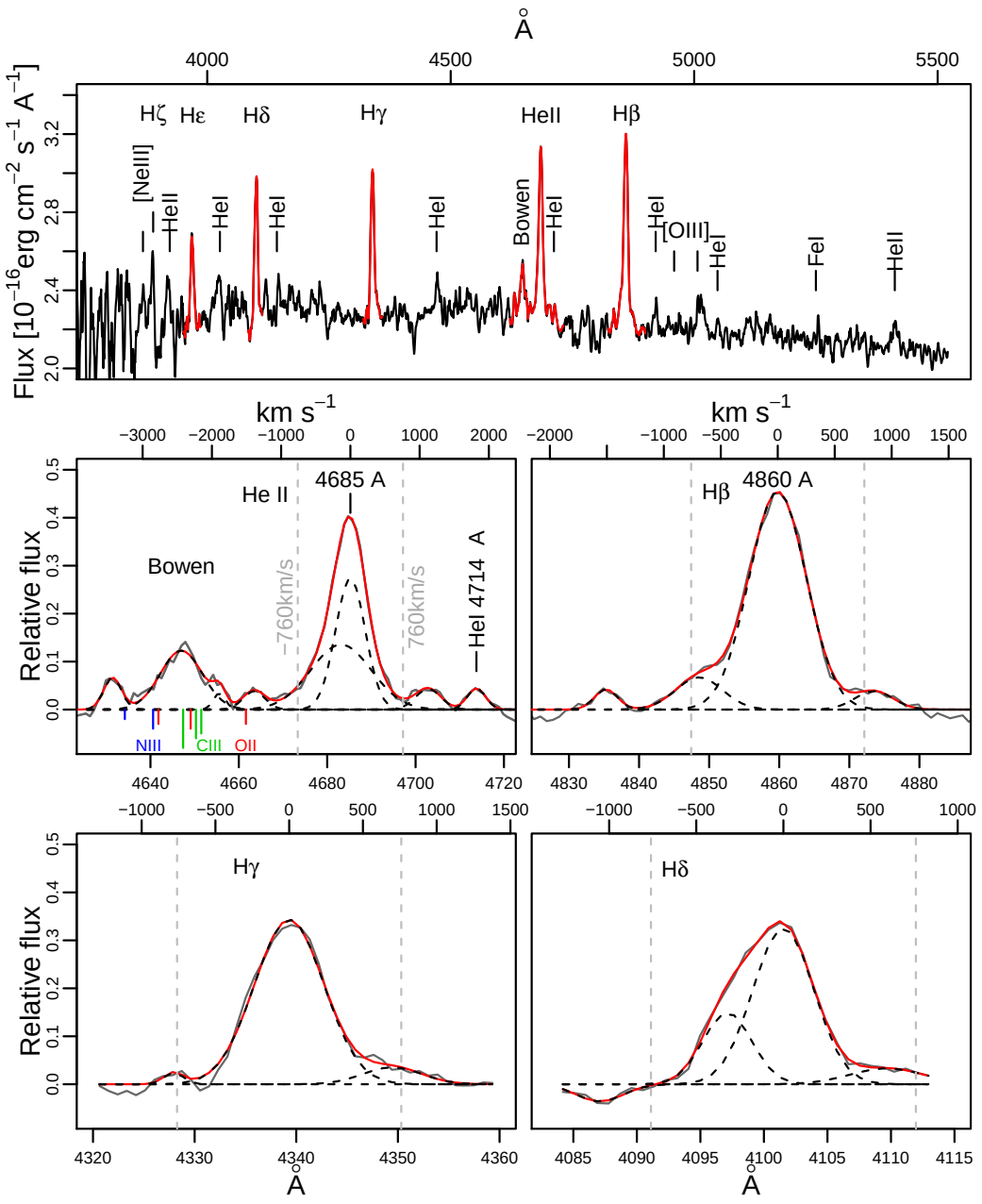

**Figure 6.** The same as in Fig. 5, but for the observations of 2015.

Because we cannot measure absolute fluxes with SALT, we are comparing here the flux normalized spectra, focusing on the emission lines. We see that both He II $\lambda$4686 and H $\beta$ appear to have become more prominent, and still saw a saddled profile that originates in an accretion disk. Three years after the outburst, unlike in most other novae the [O III] lines are still prominent and show a very complex profile with multiple components. This result indicates that the fragmentation of the shell that produces the clumps and knots observed in the shell of the previous outbursts (Shara et al. 1997) occurs at an early stage, seemingly contradicting the model of Toraskar et al. (2013) in which a Richtmyer-Meshkov instability (equivalent of a Raleigh-Taylor instability in an accelerated gas) occurs long after the outburst, causing fragmentation.

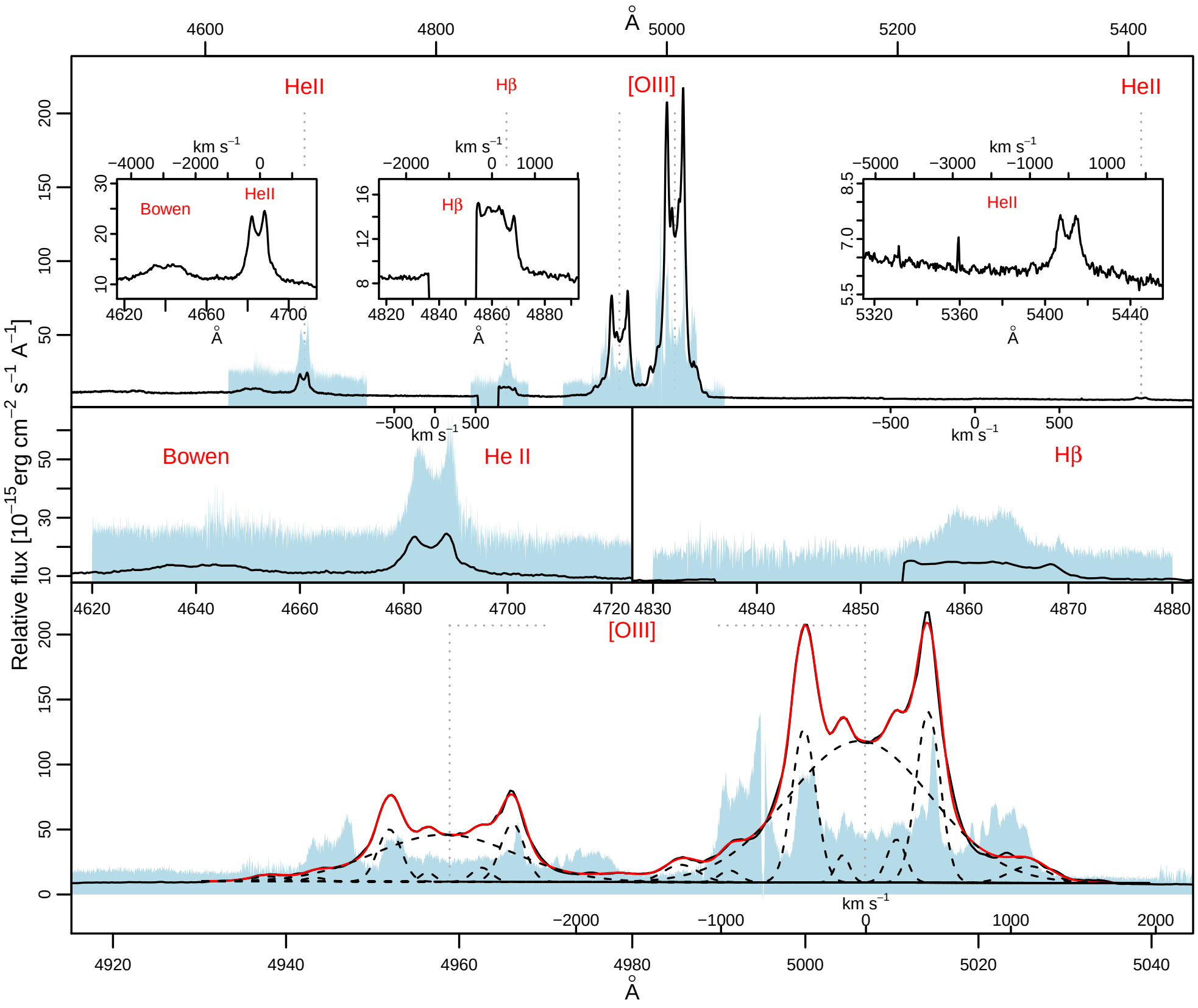

**Figure 7.** Top panel: the spectrum of T Pyx. The insets show the regions of the He II and the H$\beta$ lines. The blue shaded regions show the SALT spectrum of T Pyx obtained 2014. Bottom panel: the fit of the [O III] line of T Pyx with a number of Gaussians. The fit is shown with the red solid line and the Gaussian components are plotted with the black dashed lines.

### 5.2 The evolution of the spectrum of KT Eri

The evolving spectrum of KT Eri from outburst to quiescence is presented on Fig. 8. In this Figure we show the normalised spectra for visibility purposes. The spectrum obtained 1 year after the outburst shows prominent inverse P Cyg profiles of the Balmer lines. None of the spectra obtained at other epochs show similar profiles. These absorptions also show a prominent gradient: the absorption feature of the H$\beta$ line is stronger than that of the H$\gamma$ line. We checked that these absorptions are not a background effect. The Bowen blend was weak or absent. The ratio of the He II line to the H$\beta$ line decreased with time, indicating cooling.

### 5.3 V382 Vel

This is the oldest nova of our sample, observed twice, 13 and almost 14 years respectively after the outburst, the longest time of all the classical novae in our sample. The emission lines of V382 Vel are markedly different from the other novae. The Bowen complex is very prominent, and all the lines are very broad, with a FWHM=1800 km s$^{-1}$. This emission may arise in material that has not slowed down significantly and is still expanding, but at this stage it is more likely that we are observing a spectrum originating in a wind from the accretion disk, and not the ejecta. There is an emission complex line close to the [O III] line at 5006 Å, already observed in 2011 (Tomov et al. 2015): it is due to two lines of Fe II at 4923 and 5018 Å and should not be confused with the nebular [O III].

### 5.4 HV Cet

In this nova the Bowen complex is practically absent, but the He II line at 4686 Å is still very strong, suggesting that there is still a hot ionizing source and that the plasma in which the He II originates is too dense for the Bowen blend. The [O III] lines are not detected any more after 6 years.

### 5.5 U Sco

The SALT RSS spectra of U Sco are presented in the bottom panel of Fig. 11. The spectra do not differ from those measured by Tomov et al. (2015) in April of 2012, and they are still remarkably similar to those observed at quiescence by Johnston & Kulkarni (1992) and by Dürbeck et al. (1993). The main characteristic of these spectra is that hydrogen lines in emission are almost absent. The only H line is the H$\alpha$, which is much weaker in the first spectrum and comparable with the He II line in the second. No other Balmer line was detected in emission, which indicates remarkable hydrogen deficit in the accreted material. To our knowledge, because there are Balmer lines in emission in outburst, U Sco has never been modeled as a He nova, yet the quiescent spectrum are very intriguing in this respect.

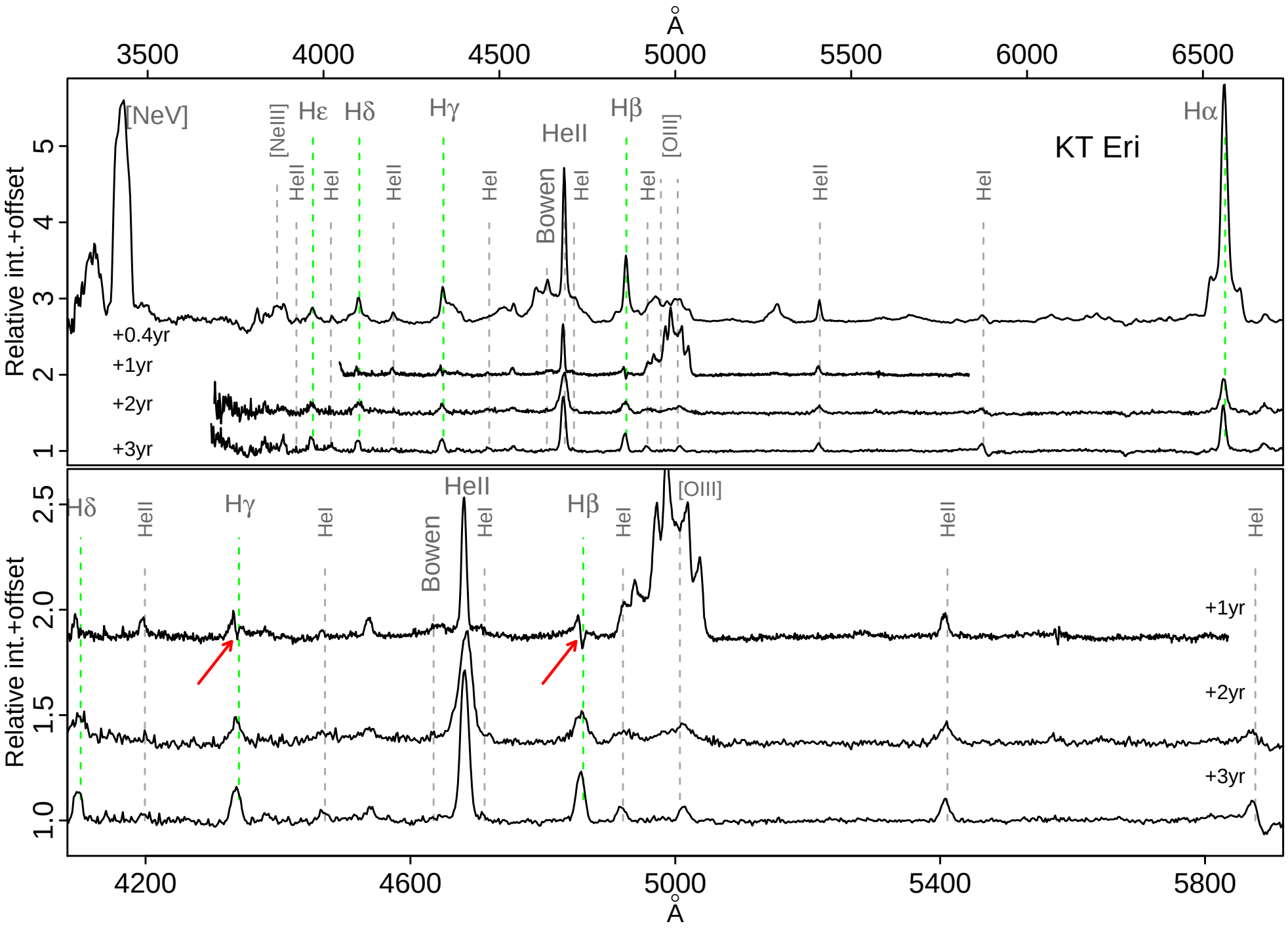

**Figure 8.** The evolution of the optical spectra of KT Eri from outburst to quiescence. The time elapsed since the outbursts is indicated. The bottom panel shows in more detail the quiescence spectra. The y-axis shows the relative intensity+arbitrary offset, chosen to the plotting purposes. Note the inverse P-Cyg profiles of the Balmer lines detected 1 year after the explosion and marked on the plot with the red arrows.

### 5.6 Some general conclusions from this group of objects

Before proceeding to discuss the spectra of the symbiotic novae, Fig. 9 shows the comparison of the quiescent spectra of 5 CNe: V4743 Sgr (published in Paper II), HV Cet, KT Eri, V2491 Cyg and V382 Vel. Measurements of EWs and fluxes are given in Table 4. Following Ringwald et al. (1996) and Tomov et al. (2015) we plotted the EWs of the H$\beta$ and He II lines and the Bowen blend versus time elapsed after the explosion The right panels of Fig. 10 show the trends in the novae of our sample. In the left panels we compare this group of mainly fast novae, to the more general nova samples of Ringwald et al. (1996) and Tomov et al. (2015). The fast novae studied in this work tend to have higher ratios of EW(He II)/EW(H$\beta$), because, having been very luminous SSS, they hosted hotter WDs from the beginning, than the larger and more diverse group of objects studied by the above authors. However, we measured generally lower values of the emission lines EWs than the larger and more diverse group of objects studied by the above authors. This may imply that the WD photosphere is cooling fast, reducing the number of photons with wavelengths shorter than the 228 Å edge, consequently contributing less to the formation of the He II $\lambda$4686 line. We also note that in the repeated observations of KT Eri, V2491 Cyg and V382 Vel, the EWs of the emission lines decreased with time, with the exception of H$\beta$ in the 2015 spectrum of V2491 Cyg, and the Bowen blend and the He II line in V382 Vel within only 10 months, suggesting a fluctuating mass accretion rate in this nova.

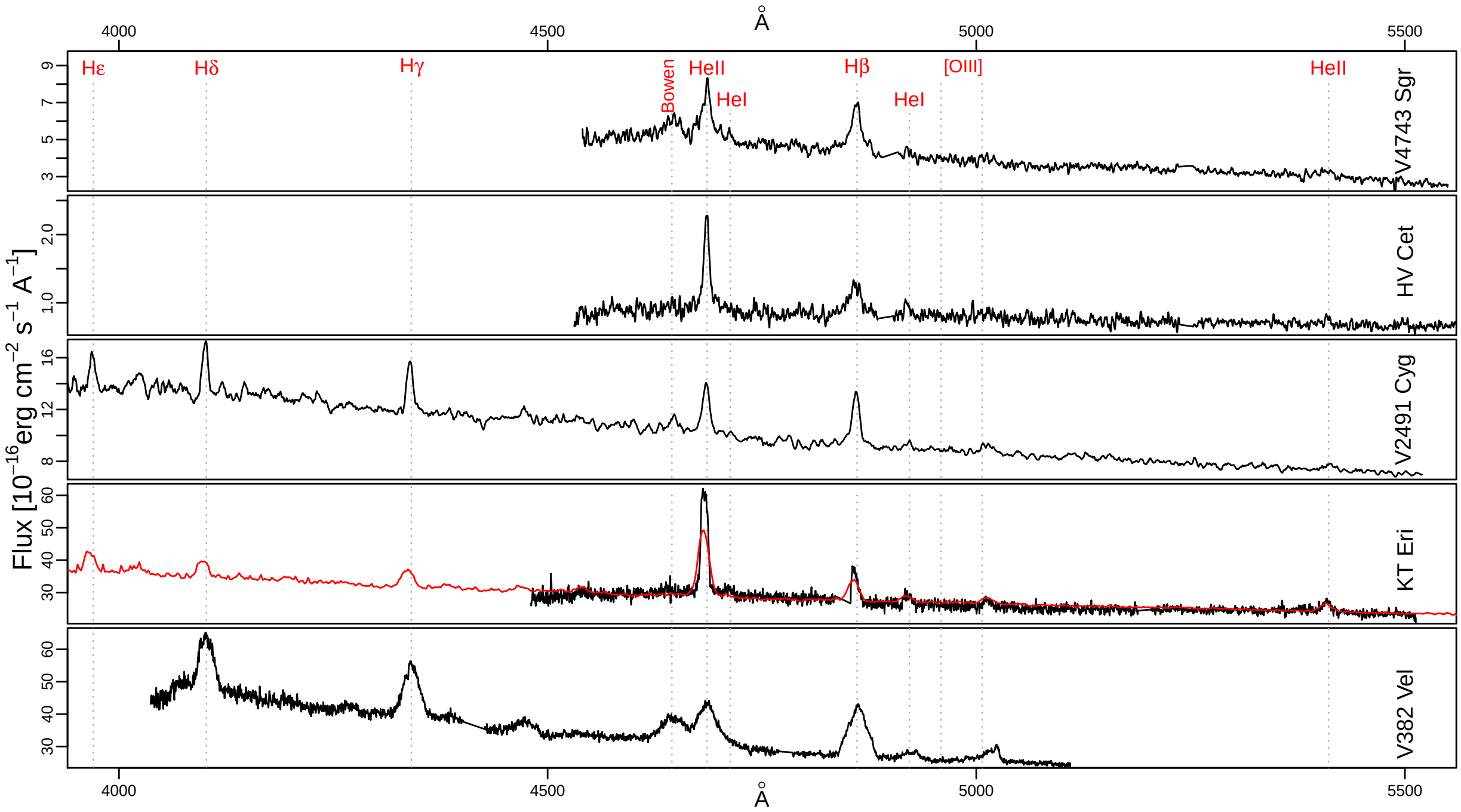

**Figure 9.** Spectra of V4743 Sgr, HV Cet, V2491 Cyg (only the last BTA spectrum), KT Eri (SALT in black, BTA in red) and V382 Vel (2013 spectrum). All the spectra were flux calibrated and de-reddened except the SALT spectrum of KT Eri. In the figure, we placed it on the y axis to match the continuum level of the BTA spectrum.

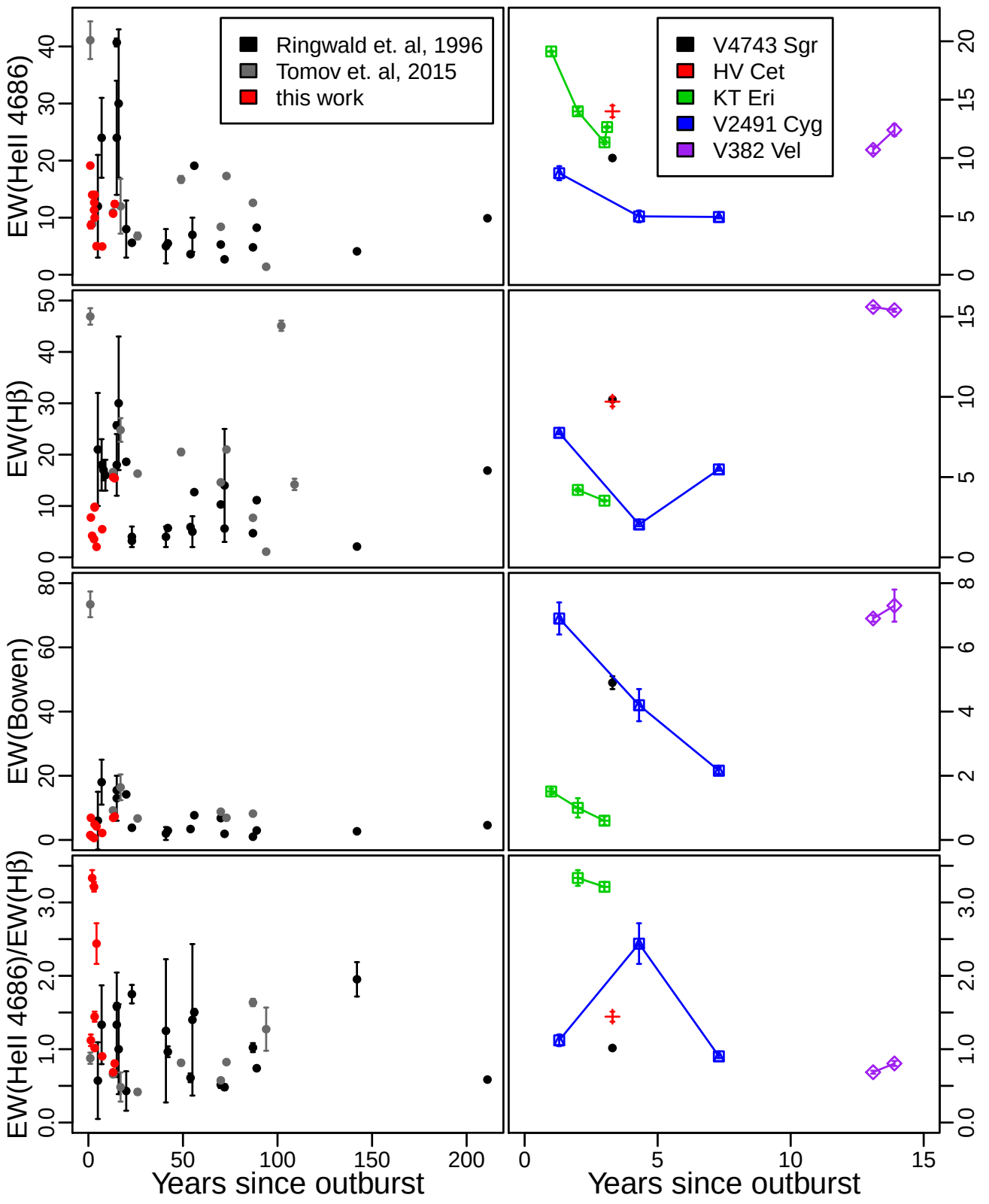

**Figure 10.** Equivalent widths of the He II, H$\beta$ and the Bowen blend versus time. The panels on the left show our data (the red points) in comparison with those of Ringwald et al. (1996) (in black) and of Tomov et al. (2015) (on grey). The right panel show only the data obtained in this work.

**Table 4.** Fluxes and EWs of the emission lines of the novae of the sample, when measurable

| Nova | yr. | Hα | | Hβ | | Hγ | | Hδ | | HeII$\lambda$4686 | | Bowen blend | |
|---|---|---|---|---|---|---|---|---|---|---|---|---|---|
| | | F[a] | EW[b] | F | EW | F | EW | F | EW | F | EW | F | EW |
| V4743 Sgr | 3.3 | | | 63±2 | 9.8±0.1 | | | | | 74±3 | 10.0±0.2 | 36±2 | 4.9±0.2 |
| | 1.3 | | | 26.8±0.4 | 7.8±0.1 | 16.4±0.3 | 4.7±0.1 | 17.9±0.3 | 5.2±0.1 | 31.2±0.7 | 8.7±0.6 | 25±2 | 6.9±0.5 |
| V2491 Cyg | 4.3 | 26.2±0.7 | 11.6±0.2 | 4.7±0.4 | 2.1±0.1 | 7.0±0.5 | 3.1±0.1 | | | 12.0±1.2 | 5.0±0.5 | 10±1 | 4.2±0.5 |
| | 7.3 | | | 12.0±0.2 | 5.5±0.1 | 7.4±0.2 | 3.3±0.1 | 6.3±0.1 | 2.8±0.1 | 11.0±0.2 | 5.0±0.1 | 4.8±0.2 | 2.2±0.1 |
| V382 Vel | 13.9 | | | | 15.4±0.1 | | 10.4±0.1 | | 7.9±0.1 | | 12.4±0.5 | | 7.3±0.5 |
| | 1 | | | | | | | | | | 19.1±0.1 | | 1.5±0.1 |
| KT Eri | 2 | | 10.4±0.1 | | 4.2±0.1 | | 2.4±0.4 | | 4.2±0.5 | | 14±0.3 | | 1±0.3 |
| | 3 | 151±1 | 10.3±0.1 | 75±1 | 3.5±0.1 | 60±10 | 2.6±0.4 | 55±10 | 2.2±0.4 | 247±3 | 11.3±0.2 | 13±12 | 0.6±0.1 |
| | 3 | | | | | | | | | 244±3 | 12.7±0.1 | | |
| HV Cet | 3.3 | | | 4.9±0.4 | 9.7±0.3 | | | | | 7.6±0.3 | 14±0.5 | | |

[a]Flux ($\times10^{-16}$ erg cm$^{-2}$ s$^{-1}$), [b]Equivalent width (Å)

## 6 THE SYMBIOTIC NOVAE

### 6.1 RS Oph

The SAAO spectra of RS Oph in two wavelength ranges, from 4000 to 4900  and from 6200 to 7000 , are presented in the top panel of Fig. 11. The dates of observations and the corresponding orbital phases are marked on the plot. The phases were calculated using the ephemerides presented in Brandi et al. (2009). The spectra show a pronounced variability of emission lines and continuum.

We could reproduce the results of the previous authors, finding that the Balmer lines have broad pedestals, wings with FWHM $\sim$200–400 km s$^{-1}$ and narrow, blue-shifted absorption features (FWHM$\sim$70–80 km s$^{-1}$) as shown in Fig. 12. The pedestals have FWZI$\sim$4000 km s$^{-1}$, larger than in previous observations of Brandi et al. (2009). While moderate changes of the central positions of the emission lines can be attributed to the orbital motion, the pedestals of the Balmer lines and the shape of the continuum changed significantly between observations on time-scales as short as a day. It should be mentioned, however, that the fast and strong variability of the pedestals of the Balmer lines, which have very asymmetric and irregular shape, affects the measurements of the central positions and the intensity of the narrow components. The double-peaked profile of the Balmer lines of RS Oph was mentioned by several authors (see Section 2.8), however, the origin of the absorption components is still a matter of debate. Considering that absorption lines were still present 2 days after the nova explosion (Patat et al. 2011), Booth et al. (2016) concluded that they are produced in material at least $5\times10^{13}$ cm away from the binary, which is probably an evaporative inflow to the poles, that also produces the pedestals of the emission lines. The narrow component of the Balmer lines in our spectra is always red-shifted with respect to the absorption one and the separation between them was $\sim$25 km s$^{-1}$ in the observations performed on 2011 and only $\sim$10 km s$^{-1}$ in 2012, so we confirm that its motion does not correlate with either the WD or the red giant, as inferred by Brandi et al. (2009).

The spectra show also numerous emission lines of Fe I and Fe II and the absorption lines of Ti I, the latter originating in the giant secondary. The He II line and the Bowen blend are weak, and their intensity changed between 2011 and 2012, probably because they were observed at different orbital phases. All our spectra show unambiguous presence of the Li I doublet, with a clear absorption features at $\lambda$6708, discovered before us by Brandi et al. (2009). However, the above author found that it originates in the secondary, and cannot be used to investigate lithium nucleosynthesis in the thermonuclear runaway.

### 6.2 V3890 Sgr

Six spectra of V3890 Sgr are shown in the middle panel of Fig. 11 and in Fig. 13. They reveal a similar H$\alpha$ line structure to that of RS Oph: emission with broad wings and a central absorption feature (see 13). The FWHM of the H$\alpha$ line lies in the range 95–105 km s$^{-1}$, the broad component has FWHM $\sim$300 km s$^{-1}$ and the absorption component — 60–80 km s$^{-1}$. The moderate changes of the intensity and central position of the H$\alpha$ line components most probably reflect the orbital motion. The spectra show the Balmer lines and lines of He I, [O I], and very weak He II at 4686 Å.

## 7 CONCLUSIONS

**1. Novae as IP candidates:** The optical photometry we presented, in which we clearly detect the beat period of the rotation and spin periods for V4743 Sgr, and we find a periodicity that may be attributed to the beat period also for V2491 Cyg, confirms the classification of the first nova as an IP and gives some credibility to the suggestion that this second nova is also an IP. Moreover, the optical spectra of V2491 Cyg do not show double peaked profiles, possibly indicating low inclination and explaining why the orbital period has never been detected. These optical spectra also present another possible diagnostic of an IP, namely a second component in the emission lines. In fact, the measurement of the beat period in optical offers a method to investigate the magnetic nature of cataclysmic variables, although detection of the spin period in X-rays has so far been considered the best indirect proof of an IP. The shrinking region that still emits supersoft X-ray flux in V4743 Sgr and V2491 Cyg is also likely to be connected with a delayed quenching of the nuclear burning at the poles, or a inhomogeneous atmosphere on the WD, two interpretations that have been proposed for recent observation of V407 Lup, another nova that is also an IP (Aydi et al. 2018).

**2. How the "weird" T Pyx is different:** Ringwald et al. (1996) found that the rate of fading of nebular lines is only dependent on the speed class of the nova. The theoretical models predict that slow novae eject larger envelopes. We expect that the time for the nebular lines to disappear is an indication of the time to disperse the ejecta in the interstellar medium. This should be a function of both the ejection velocity and mass of ejecta, which are anti-correlated in the simulations. Therefore, for slow novae we expect that the [O III] lines last for much longer periods than fast novae, that tend to have higher ejection velocities and smaller ejecta mass. Indeed, we find that T Pyx maintains much more prominent [O III] lines than the other, faster novae we observed spectroscopically.

Most interestingly, the high resolution optical spectra of T Pyx show a fragmented and complicated structure in the [O III] nebular lines, supporting a scenario in which the shell fragmentation is due to the outflow mode and not to later instabilities occurring in the shell.

**3. How long, and why, do the nova WDs remain hot?** The spectra of five CNe, presented in Fig. 9, are very similar, but V382 Vel has much broader lines than the other novae. These spectra show prominent He II lines, and the Bowen blend is also still present in all, except KT Eri. We note that Williams & Ferguson (1983) stress that "The fluorescence mechanism requires: 1) high line resonance optical depths to produce trapping of the line radiation so conversion to optical transitions can take place, 2) densities which are sufficiently low such that collisional de-excitation of the levels does not occur, 3) relatively small velocity gradients within the emitting volume to ensure the 'tuning' between the [O III] and the N III resonant transitions."

It follows that the conditions for the $\lambda$4686 He II and for the Bowen blend may be quite different, but both can originate in inner zones of the accretion disk and their relative strength with respect to H$\beta$ is an indication of high $\dot{m}$. However, because the Bowen blend requires low density, the He II line may also originate in a medium with density that is too high for the Bowen blend. The Bowen blend may have origin in the decelerated, ejected nova shell, while the He II line at this post-outburst stage is always formed in the accretion disk or accretion stream. In any case, both the ratio of Bowen to H$\beta$ and that of the He II line to H$\beta$ can trace the cooling of the WD atmosphere as it returns to quiescence; large ratios should indicate that mass transfer has resumed at high rate because of irradiation of the secondary.

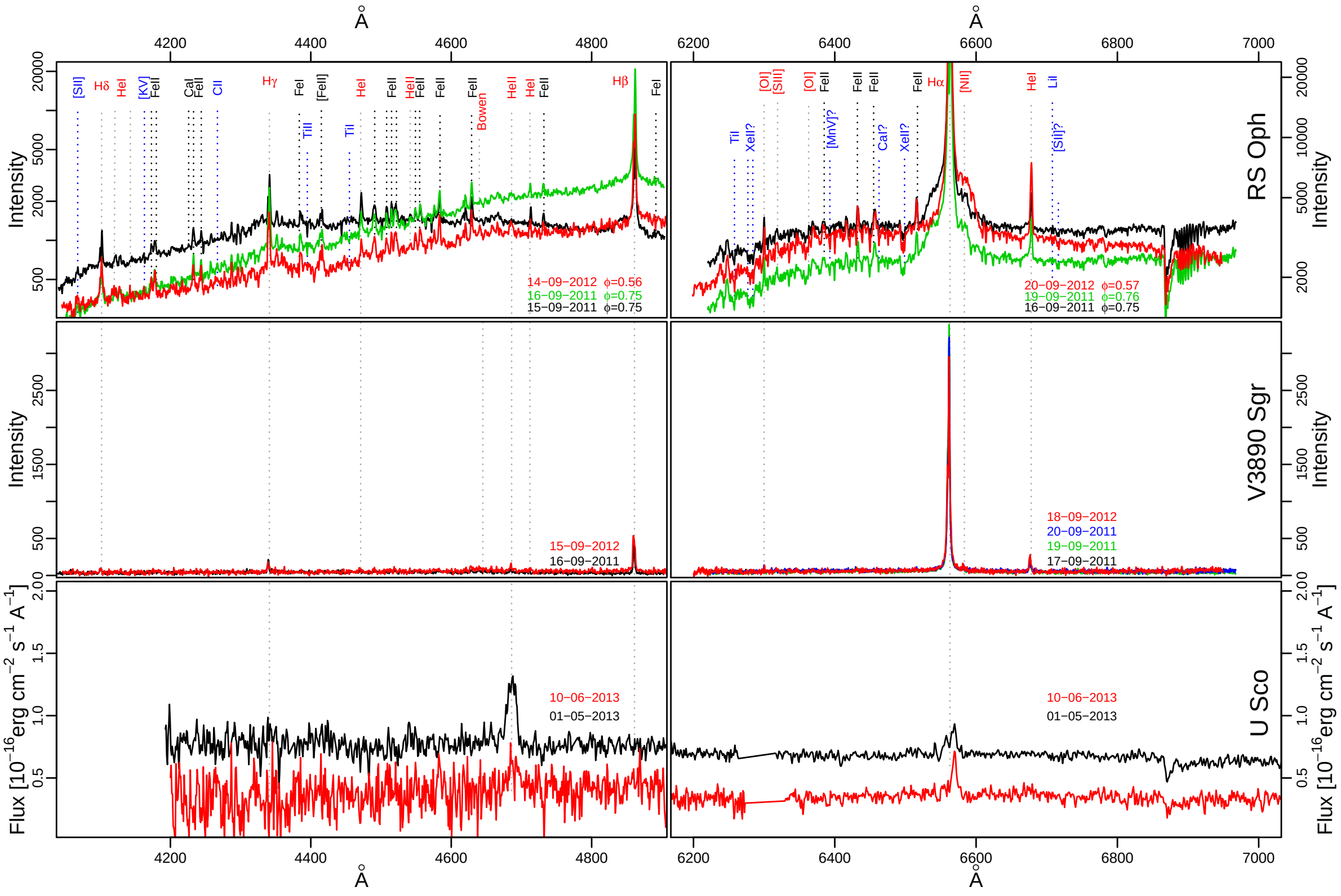

**Figure 11.** The SAAO spectra of RS Oph and V382 Vel and the SALT spectra of U Sco. The dates of the observations are marked in each panel. For RS Oph we also indicate the orbital phase.

However, after comparing our measurements also with those of Ringwald et al. (1996) and Tomov et al. (2015), we found that the equivalent widths of the H$\beta$, He II and Bowen blend and and the ratio, of intensity and especially of EW, of He II and Bowen blend to that of H$\beta$ do not seem to be correlated with post-outburst age. neither do we find a correlation with orbital period, as expected if irradiation triggers mass transfer, confirming results by Ringwald et al. (1996) and Tomov et al. (2015).

There has been speculation in the literature that short period systems have longer SSS phases by refueling the burning with enhanced mass accretion of an irradiated secondary (e.g. Greiner et al. 2003). Enhanced mass transfer in the early post-outburst phase, dependent on the orbital separation, is also predicted in the hibernation scenario (Shara et al. 1986; Kovetz et al. 1988): a period of high $\dot{m}$ would precede and follow every nova explosion, with a long inter-outburst period at very low $\dot{m}$ in between. After observational evidence discovered by Naylor et al. (1992) against the hibernation scenario, Ringwald et al. (1996), and more recently Tomov et al. (2015) found that the intensity ratio of the He II and Bowen lines to H$\beta$ does not appear to become weaker in very old novae, indicating a steady average $\dot{m}$ on long timescales. However, Ringwald et al. (1996) also discovered that several very old novae cannot be spectroscopically identified, a fact that may imply that they have stopped (or almost stopped) accreting. If hibernation occurs, we expect the ratio of the He II and Bowen lines to H$\beta$ to be large and to remain so, because of high mass transfer rate triggered after the outburst, with inverse dependence on the orbital period, because a short orbital distance allows more effective irradiation. Our data support, albeit marginally, the lack of evolution of the lines' EW ratios, but we did not find any clear correlation with orbital periods given in Table 1.

**4. The symbiotic novae:** Finally, we must notice the difference between the spectra of RS Oph and V3890 Sgr with most symbiotics that are "permanent" SSS and appear to be burning hydrogen steadily (see e.g. spectrum of SMC 3 in Orio et al. 2007). There is a symbiotic in M31 that does not have coronal lines of [Fe X] and seems to be a steady nuclear burner, yet it is a steady SSS source (Orio et al. 2017). In that system, the plasma density must be higher than 5 $\times 10^9$ cm$^{-3}$ to avoid forming coronal lines, yet the inclination of the disk and the filling factor the nebular material must be such to allow observing the SSS directly. However, RS Oph and V3890 Sgr are neither persistent SSS, nor do they emit coronal lines. Also the absence or extreme weakness of the He II at 4686 Å in RS Oph and V3890 Sgr, indicate either that the WD has completely cooled, or that there is a very high density material in the binary. Hydrogen burning is intermittent in the RNe-symbiotics. We would like to suggest a possible interpretation of this phenomenon: most likely, this is because the mass transfer rate itself is intermittent and pauses between outbursts, or in any cases it does not occur at a constant rate.

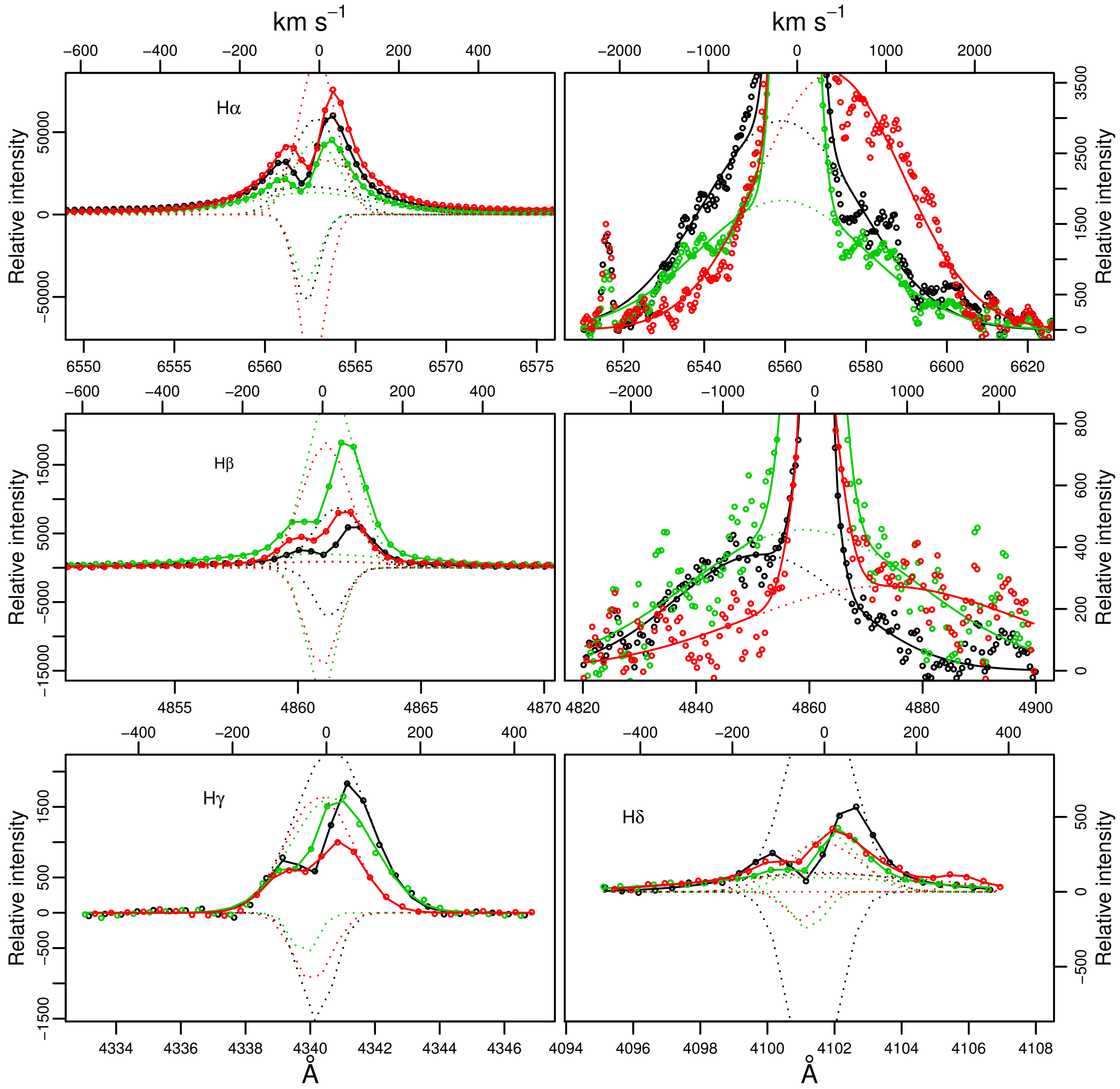

**Figure 12.** The fits to the emission lines of RS Oph. The different epochs are color-coded as in the top panel of Fig. 11. The left panels show the double-peaked emission lines and how we modeled them with one or two Gaussians for the broad emission, and one narrow Gaussian for the blue-shifted absorption. The right panels next to Hα and H$\beta$ show the broad pedestals. The dots represent the data, the solid lines the result of the fit and the dots follow the observed spectrum.

## ACKNOWLEDGEMENTS

We thank N.V. Metlova for her observations of HV Cet that we presented in the paper. This paper is based on observations made at the South African Astronomical Observatory (SAAO, A. PI Odendaal) and at the South African Large Telescope (SALT, PI M. Orio), obtained under programs: 830-4 2012-2-UW-004, 1096-5 2013-1-UW_OTH-001, 1275-6 2013-2-UW-006, 1178-7 2013-2-UW-001, 1178-7 2013-2-UW-001. E. Barsukova was funded by the Russian Foundation for Basic Research research project No.16-02-00758. A.F. Valeev is supported by the Russian Science Foundation under grant 14-50-00043. The CSS survey is funded by the National Aeronautics and Space Administration under Grant No. NNG05GF22G issued through the US Science Mission Directorate Near-Earth Objects Observations Program. The CRTS survey is supported by the U.S. National

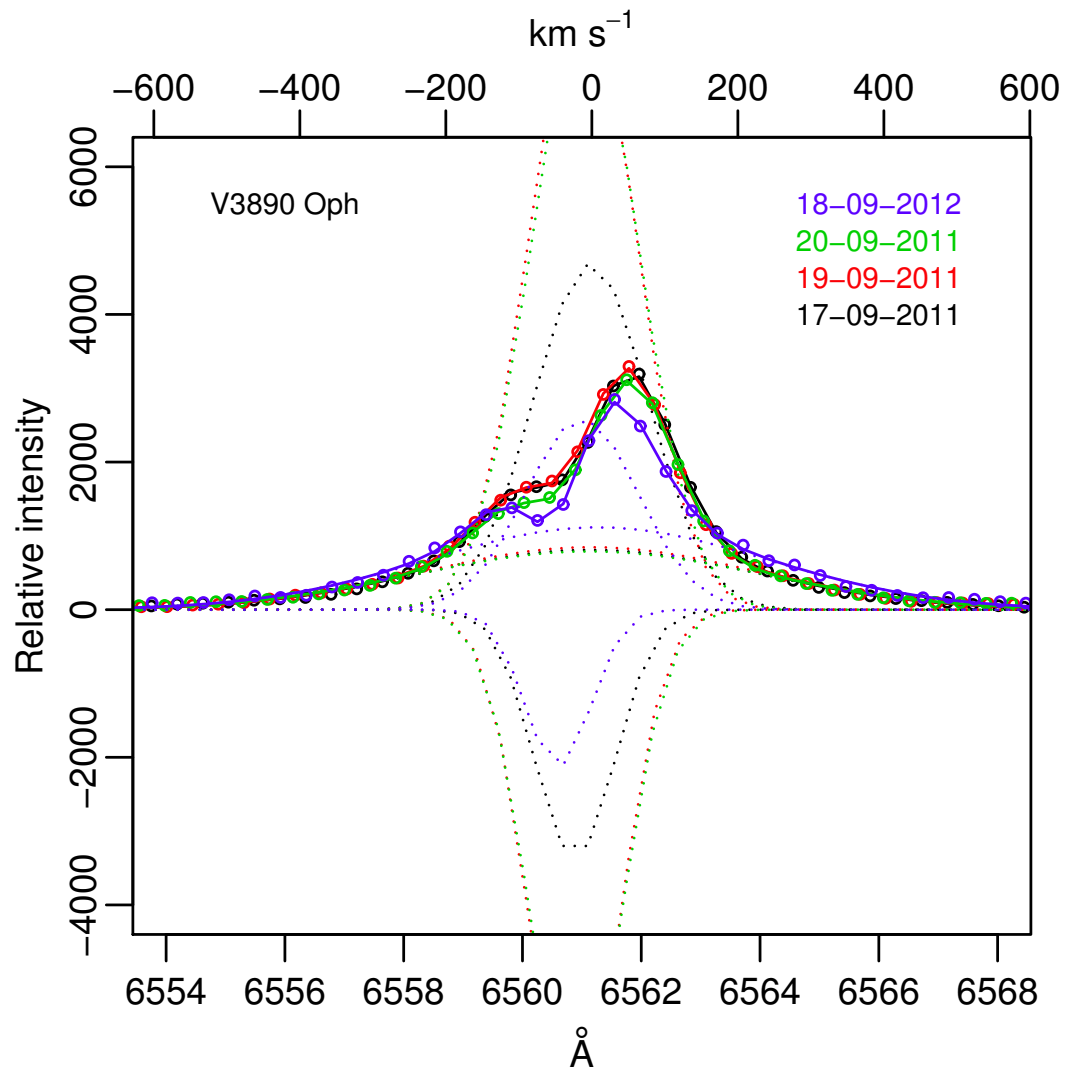

**Figure 13.** The same as in Fig. 12, for the Hα line of V3890 Sgr.

Science Foundation under grants AST-0909182 and AST-1313422. The work on V2491 Cyg and V4743 Sgr reported in this paper is included in P. Zemko's PhD thesis, funded by the CARIPARO Foundation at the University of Padova.